\documentclass[1p,sort&compress]{elsarticle}
\usepackage{amssymb}
\usepackage{amsmath}
\usepackage{amsthm}
\usepackage{amsfonts}
\usepackage{hyperref}
\usepackage{graphicx}
\usepackage{float}
\usepackage{epic}
\usepackage[section]{algorithm}
\usepackage{algpseudocode}
\usepackage{setspace}
\usepackage{multirow}
\usepackage{subcaption}
\usepackage{color}
\usepackage{booktabs}
\usepackage{dirtree}

\theoremstyle{definition}
\newtheorem{definition}{Definition}[section]

\newtheorem{proposition}{Proposition}[section]
\newtheorem{example}{Example}[section]

\numberwithin{equation}{section}

\def\x{\mathbf{x}}
\def\X{\mathbf{X}}
\def\thbf{\boldsymbol{\theta}}
\def\E{{\rm E}\,}

\begin{document}

\begin{frontmatter}

\title{Prediction based on conditional distributions of vine copulas}

\author[ubc]{Bo Chang\corref{cor1}}
\ead{bchang@stat.ubc.ca}

\author[ubc]{Harry Joe}
\ead{harry@stat.ubc.ca}

\cortext[cor1]{Corresponding author}
\address[ubc]{Department of Statistics, University of British Columbia, Vancouver, BC V6T 1Z4, Canada}

\begin{abstract}
Vine copulas are a flexible tool for multivariate non-Gaussian distributions. For data from an observational study where the explanatory variables and response variables are measured together, a proposed vine copula regression method uses regular vines and handles mixed continuous and discrete variables. This method can efficiently compute the conditional distribution of the response variable given the explanatory variables. The performance of the proposed method is evaluated on simulated data sets and a real data set. The experiments demonstrate that the vine copula regression method is superior to linear regression in making inferences with conditional heteroscedasticity. 
\end{abstract}


\end{frontmatter}

\section{Introduction}

In the context of an observational study, where the response variable $Y$
and the explanatory variables $\mathbf{X} = (X_1,\ldots,X_p)$ are
measured simultaneously, a natural approach is to fit a joint
distribution to $(X_1,\ldots,X_p,Y)$ 
assuming a random sample 
$(x_{i1},\ldots,x_{ip},y_i)$ for $i=1,\ldots,n$, and then obtain
the conditional distribution of $Y$ given $\mathbf{X}$ for making
predictions.
Observational studies are studies where researchers observe subjects and measure several variables together, and inferences of interest are relationships
among the measured variables, including the conditional distribution of
$Y$ given other variables when there is a variable $Y$ that one may want
to predict from the other variables.
In contrast, in experimental studies, the explanatory variables (treatment factors) are controlled for by researchers, and the effect of the non-random explanatory variables is then observed on the experimental units.
The inferences of interest may be different for experimental studies.

The conditional expectation $\E(Y|\X=\x)$ and conditional quantiles
$F_{Y|\X}^{-1}(p|\x)$
can be
obtained from the conditional distribution for out-of-sample
point estimates and prediction intervals.
This becomes the usual multiple regression if the joint distribution
of $(\mathbf{X},Y)$ is multivariate Gaussian.
Unlike multiple regression, the joint-distribution-based approach uses information
on the distributions of the variables and does not specify a simple
linear or polynomial equation for the conditional expectation.

When the explanatory variable is a scalar and continuous ($p=1$), the joint distribution of $(X, Y)$ can be modeled using a bivariate parametric copula family. 
\citet{bernard2015conditional} show how different copula families
can lead to quite different shapes in the conditional mean function
$\E(Y|X=x)$ and say that linearity of conditional quantiles is a pitfall
of quantile regression.
There are applications of bivariate or low-dimensions copulas for
regression in \citet{Bouye.Salmon2009,Noh.ElGhouch.ea2013}.
However, none of the previous papers link the shape of conditional
quantiles to tail properties of the copula family.

For the multivariate distribution approach to
work for moderate to large dimensions, there are two major questions to be addressed: 
\textit{(A) How to model the joint distribution of
$(X_1,\ldots,X_p,Y)$ when $p$ is not small and some $X_j$ variables
are continuous and others are discrete?
(B) How to efficiently compute the conditional distribution of $Y$ given $\mathbf{X}$?} 
For question (A),
the vine copula or pair-copula construction is a flexible tool in high-dimensional dependence
modeling \citep{Bedford.Cooke2002,Aas.Czado.ea2009,Brechmann.Czado.ea2012,dissmann2013selecting,joe2014dependence}.

The possibility of applying copulas for prediction and regression has been explored, but an algorithm is needed in general for (B) when some variables are continuous and others are discrete. \citet{parsa2011copula} use a multivariate Gaussian copula to model the joint distribution, and conditional distributions have closed-form expressions. However, Gaussian copulas do not handle tail dependence or tail asymmetry, so can lead to incorrect inferences in the joint tails. Vine copulas are used by \citet{Kraus20171,schallhorn2017d} for quantile regression, but the vine structure is restricted to a boundary class of vines called the D-vine. A general regular-vine (R-vine) copula is adopted in \citet{cooke2015vine}, for the case where the response variable and explanatory variables are continuous. \citet{Noh.ElGhouch.ea2013} use a non-parametric kernel density approach for conditional expectations, but this can run into sparsity issues as the dimension increases.

In this paper, we propose a method, called vine copula regression, 
that uses R-vines 
and handles mixed continuous and discrete variables. That is, the predictor
and response variables can be either continuous or discrete. As a result, we
have a unified approach for regression and (ordinal) classification.
The proposed approach is interpretable, and various shapes of conditional
quantiles of $y$ as a function of $\mathbf{x}$ can be obtained depending
on how pair-copulas are chosen on the edges of the vine.
Another contribution of the paper is a theoretical analysis of the asymptotic
conditional cumulative distribution function (CDF) and quantile function for vine copula regression. 
This analysis sheds light on the
flexible shapes of $\E(Y|\X=\x)$, as well as provide guidelines on
choices of bivariate copulas on
the vine to achieve different asymptotic behavior.
For example, with the approach of adding polynomial terms
to an equation in classical multiple regression, one cannot get
monotone increasing  $\E(Y|\X=\x)$ functions that flatten out for
large values of predictor variables.

The remainder of the paper is organized as follows.
\autoref{sec:vine_copulas} gives an overview of vine copulas.
\autoref{sec:vine_copula_regression} describes the model fitting procedure and the prediction algorithm given a fitted vine regression model.
\autoref{sec:theoretical_results} provides theoretical results on how the
choices of bivariate copulas in a vine affect the asymptotic
tail behaviors of the conditional CDF and quantile function.
These results are more general than those given in 
\citet{bernard2015conditional} and provide insights into the possible
tail behaviors for higher-dimensional copulas.
Sections~\ref{sec: Simulation_Study} and \ref{sec:application} present a
simulation study and applications of vine regression.
\autoref{sec:conclusion} concludes the paper.
The supplementary materials include the code and data for Sections~\ref{sec: Simulation_Study} and \ref{sec:application}.

\section{Vine copulas}
\label{sec:vine_copulas}

In this section, we provide an overview of vine copulas.
A $d$-dimensional copula $C$ is a multivariate distribution on the unit hypercube $[0,1]^d$, with all univariate margins being $U(0,1)$. 
Sklar's theorem provides a decomposition of a $d$-dimensional distribution into two parts: the marginal distributions and the associated copula~\citep{sklar1959fonctions}.
It states that for a $d$-dimensional random vector $\mathbf{Y} = (Y_1, Y_2,
\ldots, Y_d)'$ following a joint distribution $F$ with the $j$th univariate
margin $F_j$, the copula associated with $F$ is a distribution function
$C:[0,1]^d \to [0,1]$ with $U(0,1)$ margins that satisfies 
\[
F(\mathbf{y}) = C(F_1(y_1), \ldots, F_d(y_d)), \quad \mathbf{y} \in \mathbb{R}^d.
\]
If $F$ is a continuous $d$-variate distribution function, then the copula $C$ is unique. 
Otherwise $C$ is unique on the set $\operatorname{Range}(F_1) \times \cdots \times \operatorname{Range}(F_d)$.

Vine copulas use bivariate copulas as the basic building blocks along with vine
graphs to specify the dependence structure. 
Sections
\ref{sec:bivariate_copulas} and \ref{sec:vine_structures} briefly review
some results for bivariate copulas and vine copulas that are
used subsequently.

\subsection{Bivariate distributions based on copulas}
\label{sec:bivariate_copulas}

Consider a bivariate random vector $(Y_1, Y_2)$ with joint CDF 
$F_{12}(y_1, y_2)$, marginal CDFs $F_1,F_2$,
and probability density function (PDF) $f_{12}(y_1, y_2)$.
By Sklar's theorem, there exists a copula $C(u_1, u_2)$ such that
$F_{12}(y_1, y_2) = C(F_1(y_1), F_2(y_2))$.

If $C$ is absolutely continuous, then its density is
$c(u_1, u_2) = \partial^2 C(u_1, u_2)/\partial u_1 \partial u_2$.
The set of  conditional CDFs given $U_2=u_2$ is 
$$C_{1|2}(u_1 | u_2) :=
\mathbb{P}(U_1 \leq u_1 | U_2 = u_2) 
= \partial C(u_1, u_2)/\partial u_2.$$
The conditional quantile function $C_{1|2}^{-1}(\cdot|u_2)$ is the inverse
function of $C_{1|2}(\cdot|u_2)$. 
The other set of conditional CDFs $C_{2|1}(\cdot|u_1)$ and quantile functions
$C_{2|1}^{-1}(\cdot|u_1)$ can be defined in a similar fashion.

Let $f_1$, $f_2$ and $f_{12}$ be the density functions of $Y_1$, $Y_2$ and $(Y_1, Y_2)$ respectively, with respect to Lebesgue measure for continuous random variables or counting measure for discrete ones.
Next is a result from \citet{Stoeber.Hong.ea2015} and
Section 3.9.5 of \citet{joe2014dependence}.
The joint density function $f_{12}$ can be decomposed as follows, 
\begin{equation}
\label{eq: bivariate_density_c_tilde}
 f_{12}(y_1, y_2) = \tilde{c}(y_1, y_2) f_1(y_1) f_2(y_2),
\end{equation}
where $\tilde{c}$ is defined below.
If $Y_j$ is discrete, we define
$F_j(y_j^-) := 
\mathbb{P}(Y_j < y_j)
= \lim_{t \uparrow y_j} F_j(t)$.

\begin{itemize}
    \item 
    If both $Y_1$ and $Y_2$ are continuous random variables, then 
    $\tilde{c}(y_1, y_2) := c(F_1(y_1), F_2(y_2))$.

    \item 
    If $Y_1$ is a discrete random variable and $Y_2$ is continuous, then 
    \begin{align*}
        f_{12}(y_1, y_2) 
        &= \frac{\partial }{\partial y_2} \left[F_{12}(y_1, y_2) - F_{12}(y_1^-, y_2) \right] \\    
        &= \left[C_{1|2}(F_1(y_1) | F_2(y_2)) - C_{1|2}(F_1(y_1^-) | F_2(y_2))     \right] f_2(y_2).
    \end{align*}
    In this case, $$\tilde{c}(y_1, y_2) := \left[C_{1|2}(F_1(y_1) | F_2(y_2)) - C_{1|2}(F_1(y_1^-) | F_2(y_2))     \right] / f_1(y_1).$$

    \item 
    If $Y_1$ is a continuous random variable  and $Y_2$ is discrete, then 
    $$\tilde{c}(y_1, y_2) := \big[C_{2|1}(F_2(y_2) | F_1(y_1)) - C_{2|1}(F_2(y_2^-) | F_1(y_1))     \big] / f_2(y_2).$$

    \item 
    If both $Y_1$ and $Y_2$ are discrete random variables, then the density of $(Y_1, Y_2)$ is:
    \begin{multline*}
        f_{12}(y_1, y_2)
        = C(F_1(y_1), F_2(y_2)) - C(F_1(y_1^-), F_2(y_2)) \\
         - C(F_1(y_1), F_2(y_2^-)) + C(F_1(y_1^-), F_2(y_2^-)).
    \end{multline*}
    In this case, 
    \begin{multline*}
        \tilde{c}(y_1, y_2) := \big[C(F_1(y_1), F_2(y_2)) - C(F_1(y_1^-), F_2(y_2))  \\
        -C(F_1(y_1), F_2(y_2^-)) + C(F_1(y_1^-), F_2(y_2^-))     \big] / \big[        f_1(y_1)        f_2(y_2)\big].
    \end{multline*}
\end{itemize}

\subsection{Vine structures}
\label{sec:vine_structures}

A regular vine (R-vine) in $d$ variables is a nested set of 
$d-1$ trees where the edges in the first tree are the nodes of the second tree, the edges of the second tree are the nodes of the third tree, etc. 
Vines and truncated vines provide a flexible approach to summarizing
dependence in a multivariate distribution with edges in the first
tree representing pairwise dependence and edges in subsequent trees 
representing conditional dependence. Vines extend Markov trees to allow for
conditional dependence. A multivariate Gaussian distribution can be 
represented through vines when parameters on the edges of the vine
are correlations in the first tree and partial correlations in subsequent trees;
for tree $\ell$ ($2\le\ell<d$), the partial correlations are conditioned on
$\ell-1$ variables.

In general,
the first tree represents $d$ variables as nodes and bivariate dependence of $d-1$ pairs of variables as edges. 
The second tree represents conditional dependence of $d-2$ pairs of
variables conditioning on another variable; nodes are the edges
in tree 1, and a pair of nodes could be connected if
there is a common variable in the pair.
The third tree represents conditional dependence of $d-3$ pairs of
variables conditioning on two other variables; nodes are the edges
in tree 2, and a pair of nodes could be connected if
there are two common conditioning variables in the pair.
This continues until tree $d-1$ has one edge that represents the
conditional dependence of two variables conditioning on the
remaining $d-2$ variables.

A formal definition, from \citet{Bedford.Cooke2002}, is as follows.
\begin{definition}
(Regular vine)
$\mathcal{V}$ is a regular vine on $d$ elements, with $E(\mathcal{V}) = \bigcup_{i = 1}^{d-1} E(T_i)$ denoting the set of edges of $\mathcal{V}$, if

\begin{enumerate}
\item
$\mathcal{V} = (T_1, \ldots, T_{d-1})$ [consists of $d-1$ trees];

\item
$T_1$ is a connected tree with nodes $N(T_1) = \{1, 2, \ldots, d\}$, and edges $E(T_1)$,
$T_{\ell}$
is a tree with nodes $N(T_{\ell}) = E(T_{\ell - 1})$ [edges in a tree becomes nodes in the next tree];

\item
(proximity) for $\ell = 2, \ldots, d -1$, for $\{n_1, n_2\} \in E(T_{\ell})$, $\# (n_1 \triangle n_2) = 2$, where $\triangle$ denotes symmetric difference and $\#$ denotes cardinality [nodes joined in an edge differ by two elements].

\end{enumerate}
\end{definition}

To get a vine copula or pair-copula construction,
for each edge $[jk|S] \in E(\mathcal{V})$ in the vine, there is a bivariate copula $C_{jk;S}$ associated with it.
Let $\tilde{c}_{jk;S}(\cdot;\mathbf{y}_S)$ be as defined in 
\autoref{sec:bivariate_copulas} for $C_{jk;S}(\cdot;\mathbf{y}_S)$ 
when the conditioning value is $\mathbf{y}_S$, and
let $C_{j|k;S}(a|b;\mathbf{y}_S)=\partial C_{jk;S}(a,b;\mathbf{y}_S)/\partial b$
and $C_{k|j;S}(b|a;\mathbf{y}_S)=\partial C_{jk;S}(a,b;\mathbf{y}_S)/\partial a$.
$C_{j|k;S}$ and $C_{k|j;S}$ are the conditional CDFs
of the copula $C_{jk;S}$.
The joint density of $(Y_1, \ldots, Y_d)$ can be decomposed according to the vine structure $\mathcal{V}$.
\begin{equation}
\label{eq: R-vine density decomposition}
f_{1:d} (y_1, \ldots, y_d)=
\prod_{i=1}^{d} f_i(y_i) \cdot 
\prod_{[jk|S] \in E(\mathcal{V})}
\tilde{c}_{jk;S}( y_j, y_k; \mathbf{y}_S).
\end{equation}
The above representation for the case of absolutely continuous 
random variables is derived in \citet{Bedford.Cooke2001}; its
extension to include some discrete variables is in Section 3.9.5 of
\citet{joe2014dependence}.
For simplicity of notation, we denote 
$F_{j|S}^+ = F_{j|S}(y_j | \mathbf{y}_S)$ and
$F_{j|S}^- = \lim_{t \uparrow y_j} F_{j|S}(t | \mathbf{y}_S)$.
If it is assumed that the copulas on edges of trees 2 to $d-1$ do not
depend on the values of the conditioning values,
then $c_{jk; S}$ and $\tilde{c}_{jk;S}$ in 
\eqref{eq: R-vine density decomposition}
do not depend on $\mathbf{y}_S$;
i.e., 
$c_{jk; S}(\cdot) = c_{jk; S}(\cdot; \mathbf{y}_S)$
and
$\tilde{c}_{jk; S}(\cdot) = \tilde{c}_{jk; S}(\cdot; \mathbf{y}_S)$.
This is called the \textit{simplifying assumption}.
With the simplifying assumption, we have the following definition of $\tilde{c}_{jk; S}$.

\begin{itemize}
\item 
If  $Y_j$ and $Y_k$ are  both continuous, then 
$\tilde{c}_{jk;S}( y_j, y_k) := c_{jk; S}(F_{j|S}^+, F_{k|S}^+)$.

\item
If  $Y_j$ is continuous and $Y_k$ is discrete, then 
\begin{equation*}
    \tilde{c}_{jk;S}(y_1, y_k) := \big[C_{k|j;S}(F_{k|S}^+ | F_{j|S}^+) 
    - C_{k|j;S}(F_{k|S}^- | F_{j|S}^+)     \big] / f_{k|S}(y_k | \mathbf{y}_S).
\end{equation*}

\item
If  $Y_j$ is discrete and $Y_k$ is continuous, then 
\begin{equation*}
    \tilde{c}_{jk;S}(y_j, y_k) := \big[C_{j|k;S}(F_{j|S}^+ | F_{k|S}^+) 
    - C_{j|k;S}(F_{j|S}^- |  F_{k|S}^+)     \big] / f_{j|S}(y_j | \mathbf{y}_S).
\end{equation*}

\item
If  $Y_j$ and $Y_k$ are  both discrete, then 
\begin{multline*}
    \tilde{c}_{jk;S}(y_j, y_k) := \big[C_{jk;S}(F_{j|S}^+, F_{k|S}^+) - C_{jk;S}(F_{j|S}^-, F_{k|S}^+)  
    \\
    -C_{jk;S}(F_{j|S}^+, F_{k|S}^-) + C_{jk;S}(F_{j|S}^-, F_{k|S}^-)     \big] 
    / \big[        f_{j|S}(y_j | \mathbf{y}_S)        f_{k|S}(y_k | \mathbf{y}_S)\big].
\end{multline*}
\end{itemize}

A $t$-truncated vine copula results 
if the copulas for trees $T_{t+1}, \ldots, T_{d-1}$ are all independence copulas, representing conditional independencies.

\section{Vine copula regression}
\label{sec:vine_copula_regression}

Consider the data structure for multiple regression with $p$ explanatory variables $x_1, \ldots, x_p$ and response variable $y$ as a sample of size $n$; the data are $(x_{i1}, \ldots, x_{ip}, y_i)$ for $i=1, \ldots, n$,
considered as independent realizations of a random vector
$(X_1,\ldots,X_p,Y)$.
If these data are considered as a sample in an observational study, 
then a natural approach is to fit a joint multivariate density to the variables $x_1, \ldots, x_p, y$. 
This can be done using a flexible, parametric vine copula.

Researchers have applied D-vine copulas to quantile regressions~\citep{Kraus20171,schallhorn2017d}.
Their approach is to sequentially constructs a D-vine: it first links the predictor with the strongest 
dependence to $y$; then a second variable with
strongest conditional dependence to $y$ given first predictor;
this procedure continues until an information criterion stops improving.
This structure learning algorithm is similar to the forward selection in multiple regression and easiest
to handle with a D-vine. 
However, it is known that forward selection does not usually produce an optimal solution.
Compared to the existing D-vine-based methods, our proposed algorithm uses R-vines and is more flexible.

Furthermore, \citet{schallhorn2017d} propose a method based
on continuous convolution to handle
discrete variables, and estimate the vine copula non-parametrically.
When variables are all monotonically related, the parametric approach
that we are using can be simpler for interpretations and check
monotonicity of conditional quantiles.

The remainder of this section is organized as follows.
\autoref{sec:model_fitting} introduces the model fitting and assessment procedure.
\autoref{sec:prediction} describes an algorithm that calculates the conditional CDF of the response variable of a new observation, given a fitted vine copula regression model.
The conditional CDF can be further used to calculate the conditional mean and quantile for regression problems, and the conditional probability mass function (PMF) for classification problems.

\subsection{Model fitting and assessment}
\label{sec:model_fitting}

Due to the decomposition of a joint distribution to univariate marginal distributions and a dependence structure among variables, a two-stage estimation procedure can be adopted. Suppose the observed data are 
$(z_{i1}, z_{i2}, \ldots, z_{id})=(x_{i1}, \ldots, x_{ip}, y_i)$, 
for $i=1, \ldots, n$ with $d=p+1$.

\begin{enumerate}
\item Estimate the univariate marginal distributions $\hat F_j$, for $j=1, \ldots, d$, using parametric or non-parametric methods. The corresponding u-scores are obtained by applying the probability integral transform: $\hat u_{ij} = 
\hat{F}_j(z_{ij})
$. 

\item Fit a vine copula on the u-scores. 
There are two components: vine structure and bivariate copulas. 
\autoref{sec: Vine_Structure_Learning} discusses how to choose a vine structure, and \autoref{sec: Bivariate_Copula_Selection} presents a bivariate copula selection procedure.

\item Compute some conditional quantiles, with some predictors fixed
and others varying, to check if the monotonicity properties are
interpretable.
\end{enumerate}

\subsubsection{Vine structure learning}
\label{sec: Vine_Structure_Learning}

In this section, we introduce methods for learning or choosing truncated R-vine structures.
From \citet{Kurowicka.Joe2011}, the total number of (untruncated) R-vines
in $d$ variables is $2^{(d-3)(d-2)}(d!/2)$.
When $d$ is small, it is possible to enumerate all $2^{(d-3)(d-2)}(d!/2)$ vines and find the best $\ell$-truncated R-vine based on some objective
functions such as those in Section 6.17 of \citet{joe2014dependence}.
However, this is only feasible for $d \leq 8$ in practice.
Greedy algorithms \citep{dissmann2013selecting} and metaheuristic algorithms \citep{brechmann2014parsimonious} are commonly adopted to find a locally optimal $\ell$-truncated vine.
The development of vine structure learning algorithms is an active research topic; various algorithms are proposed based on different heuristics.
However, no heuristic method can be expected to be universally the best.

The goal of vine copula regression is to find the conditional distribution of the response variable, given the explanatory variables.
In general, to calculate the conditional distribution from the joint distribution specified by a vine copula, computationally intensive multidimensional numerical integration is required.
This could be avoided if we enforce a constraint on the vine structure
such that
the node containing the response variable as a conditioned variable is always a leaf node in $T_{\ell}, \ell = 1, \ldots, d-1$.
When this constraint is satisfied, Algorithm \ref{alg: rvinepcond mix} computes the conditional CDF without numerical integration.

\begin{figure}
\centering
\begin{subfigure}[b]{0.3\textwidth}
    \centering
    \includegraphics[width=\textwidth]{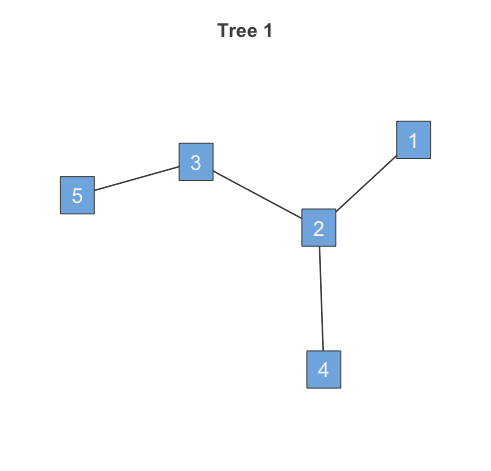}
\end{subfigure}%
\qquad
\begin{subfigure}[b]{0.3\textwidth}
    \centering
    \includegraphics[width=\textwidth]{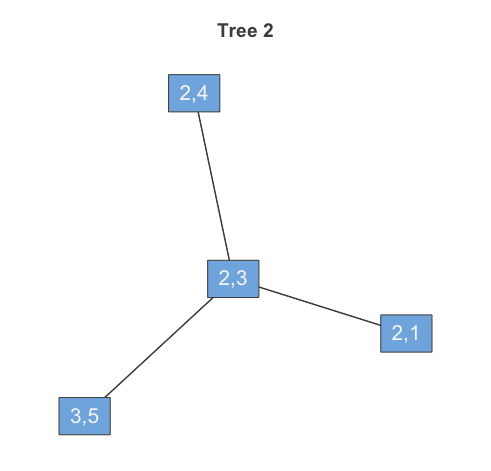}
\end{subfigure}
\caption{First two trees $T_1$ and $T_2$ of a vine $\mathcal{V}$. 
The node set and edge set of $T_1$ are $N(T_1) = \{1, 2, 3, 4, 5\}$ and $E(T_1) = \{[12], [23], [24], [35]\}$.
The node set and edge set of $T_2$ are $N(T_2) = E(T_1) = \{[12], [23], [24], [35]\}$ and $E(T_2) = \{[13|2], [25|3], [34|2]\}$.}
\label{fig: 2-trunc_vine_example}
\end{figure}

\begin{figure}
\centering
\begin{subfigure}[b]{0.3\textwidth}
    \centering
    \includegraphics[width=\textwidth]{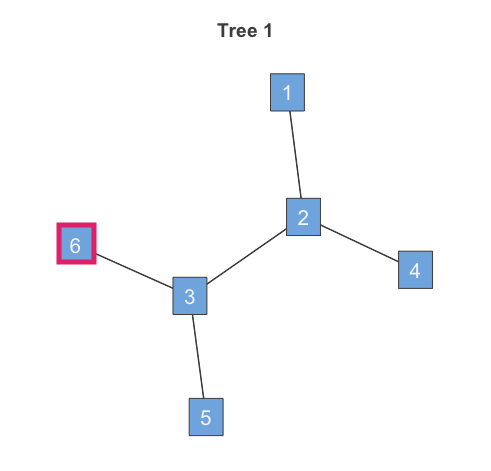}
\end{subfigure}%
\qquad
\begin{subfigure}[b]{0.3\textwidth}
    \centering
    \includegraphics[width=\textwidth]{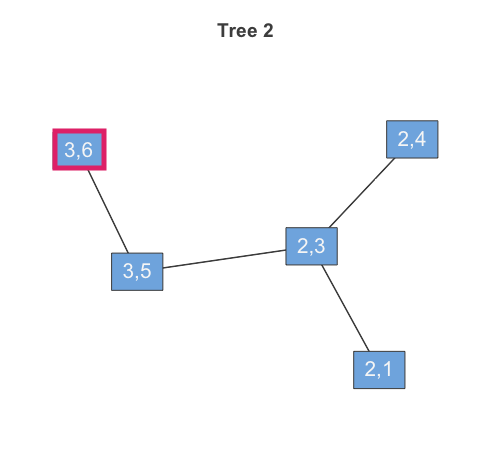}
\end{subfigure}
\caption{Adding a response variable to the R-vine of the explanatory variables. In this example, variables 1 to 5 represent the explanatory variables and variable 6 represents the response variable. 
The newly added nodes are highlighted.}
\label{fig: adding a response variable}
\end{figure}

To construct a truncated R-vine that satisfies the constraint, we can first find a locally optimal $t$-truncated R-vine using the explanatory variables $x_1, \ldots, x_p$. Then from level 1 to level $t$, the response variable $y$ is sequentially linked to the node that satisfies the proximity condition and has the largest absolute (normal scores) correlation with $y$.
The idea of extending an existing R-vine is also explored by
\citet{bauer2016pair} for the construction of non-Gaussian conditional
independence tests.
Figures~\ref{fig: 2-trunc_vine_example} and \ref{fig: adding a response variable} demonstrate how to add a response variable to the R-vine of the explanatory variables, after each variables has
been transformed to standard normal $N(0,1)$. 
Given a $2$-truncated R-vine $\mathcal{V} = (T_1, T_2)$ in \autoref{fig: 2-trunc_vine_example} with $N(T_1) = \{1, \ldots, 5\}$, $E(T_1) = N(T_2) = \{[12], [23], [24], [35]\}$, $E(T_2) = \{[13|2], [25|3], [34|2]\}$. 
Suppose the response variable is indexed by $6$. 
The first step is to find the node that has the largest absolute correlation, i.e. $\arg\max_{1 \leq i \leq 6} |\rho_{i6}|$.
Assume $\rho_{36}$ is the largest, then node $3$ and node $6$ are linked: $N(T_1') = N(T_1) \cup \{6\}$, $E(T_1') = E(T_1) \cup \{[36]\}$. 
At level 2, according to the proximity condition, node $[36]$ can be linked to either $[23]$ or $[35]$. So we compare $\rho_{26;3}$ with $\rho_{56;3}$. 
If we assume $|\rho_{56;3}| > |\rho_{26;3}|$, then $E(T_2') = E(T_2) \cup \{[56|3]\}$.
So the new $2$-truncated R-vine is $\mathcal{V}' = (T_1', T_2')$, as shown in
\autoref{fig: adding a response variable}.

\subsubsection{Bivariate copula selection}
\label{sec: Bivariate_Copula_Selection}

After fitting the univariate margins and deciding on the vine structure, parametric bivariate copulas can be fitted sequentially
from tree 1, tree 2, etc.
The results in Section \ref{sec:theoretical_results} can provide
guidelines of choices of bivariate copula families in order to 
match the expected behavior of conditional quantile functions in the
extremes of the predictor space.
With the simplifying assumption and parametric copula families, 
the log-likelihood of the bivariate copula $C_{j,k;S}$ on edge
$[jk|S]\in E(\mathcal{V})$, is 
\[
\ell_{jk;S}({\boldsymbol\theta}_{jk}) = \sum_{i=1}^n \log \big( \tilde{c}_{jk;S}( 
z_{ij}, z_{ik};
{\boldsymbol\theta}_{jk})\big).
\]
Commonly used model selection criteria include
Akaike information criterion (AIC) and Bayesian information criterion (BIC):
\begin{align*}
\mathrm{AIC}_{jk;S}(\thbf_{jk}) &= -2\ell_{jk;S}(\thbf_{jk}) + 2|\thbf_{jk}|, 
\\
\mathrm{BIC}_{jk;S}(\thbf_{jk}) &= -2\ell_{jk;S}(\thbf_{jk}) + \log(n)|\thbf_{jk}|,
\end{align*}
where $|\thbf_{jk}|$ refers to the number of copula parameters 
in $c_{jk;S}$.
For each candidate bivariate copula family on an edge, we first find the parameters that maximize the log-likelihood $\hat{\thbf}_{\mathrm{MLE}}$. 
Then the copula family with the lowest AIC or BIC is selected.
When all the variables are continuous, this approach of selecting the bivariate copula selection is the standard approach in \texttt{VineCopula} \citep{schepsmeier2018vinecopula} and has been initially proposed and investigated by \citet{brechmann2010truncated}.

\subsection{Prediction}
\label{sec:prediction}

This section describes how to predict the conditional distribution of the response variable of a new observation, given a fitted vine copula regression model.
We first present an algorithm that computes the conditional CDF of the response variable.
If the response variable is continuous, the conditional quantile and mean can be calculated by inverting the conditional CDF and integrating the quantile function.
If the response variable is discrete, the conditional PMF can be easily derived from the conditional CDF via finite difference.

Based on ideas of the algorithms in Chapter 6 of \citet{joe2014dependence},
Algorithm \ref{alg: rvinepcond mix} can be applied to an R-vine
with mixed continuous and discrete variables.
The idea is that, given the structural constraint on the vine structure described in \autoref{sec: Vine_Structure_Learning}, conditional distributions are sequentially computed according to the vine structure, and the conditional distribution of the response variable given all the explanatory variables is obtained in the end.
The input is a vine copula regression model with a vine array $A = (a_{kj})$, a vector of new explanatory variables $\mathbf{x} = (x_1, \ldots, x_d)'$, and a percentile $u \in (0, 1)$.
The vine array is an efficient and compact way to represent a vine
structure; see \ref{sec:vine_array_representation}
or \citet{Kurowicka.Joe2011} or \citet{joe2014dependence}.
The R-vine matrices in the \texttt{VineCopula} package \citep{schepsmeier2018vinecopula} are the vine arrays with backward indexing
of rows and columns.
The algorithm returns the conditional CDF of the response variable given the explanatory variables evaluated at $u$, that is, $p(u|\mathbf{x}) := \mathbb{P}(F_Y(Y) \leq u | \mathbf{X} = \mathbf{x})$.
It calculates the conditional distributions 
$C_{j | a_{\ell j} ; a_{1j}, \ldots, a_{\ell-1, j}}$ and $C_{a_{\ell j} | j; a_{1j}, \ldots, a_{\ell-1, j}}$ for $\ell = 1, \dots, n_{\mathrm{trunc}}$ and $j = \ell+1, \ldots, d$, where $n_{\mathrm{trunc}}$ is the truncation level of the vine copula.
For discrete variables, both the left-sided and right-sided limits of the conditional CDF are retained.
In the end, $C_{d | a_{d-1, d}; a_{1d}, \ldots, a_{d-2,d}}$ is returned.

If the response variable $Y$ is continuous, then the conditional mean and conditional quantile can be calculated using $p(\cdot|\mathbf{x})$: the $\alpha$-quantile is $F_Y^{-1}(p^{-1}(\alpha|\mathbf{x}))$, and the conditional mean is $$\mathrm{E}(Y | \mathbf{X} = \mathbf{x}) = \int_0^1 F_Y^{-1}(p^{-1}(\alpha | \mathbf{x})) \mathrm{d} \alpha,$$
where $p^{-1}(\cdot | \mathbf{x})$ is calculated using the secant method, and the numerical integration is computed using Monte Carlo methods or numerical quadrature.
If the response variable $Y$ is ordinal, then it is a classification problem; we only need to focus on the support of $Y$. 
The conditional CDF is fully specified by $p(F_Y(y)) = \mathbb{P}(Y \leq y | \mathbf{X} = \mathbf{x})$, where $y \in \{k: \mathbb{P}(Y = k) > 0\}$.

If the response variable $Y$ is nominal, then the proposed method does not apply. 
An alternative vine-copula-based method is to fit a vine copula model for each class separately and use the Bayes' theorem to predict the class label. 
Specifically, for samples in class $Y=k$, we fit a vine copula density $\hat{f}_{\mathbf{X}|Y}(\mathbf{x}|k)$. Let $\hat{\pi}_k$ be the proportion of samples in class $k$ in the training set. According the Bayes' theorem, the predicted probability that a sample belongs to class $k$ is
\begin{equation*}
\hat{f}_{Y|\mathbf{X}}(k | \mathbf{x}) = \frac{\hat{\pi}_k \hat{f}_{\mathbf{X}|Y}(\mathbf{x}|k)}{\sum_{j} \hat{\pi}_j \hat{f}_{\mathbf{X}|Y}(\mathbf{x}|j)}.
\end{equation*}
The classification rule has been utilized in \citet{nagler2016evading} in an example involving vines with nonparametric pair copula estimation using kernels.
Since the distribution of predictors is modeled separately for each class, this alternative method is more flexible but has a high computational cost, especially when the number of classes is large.

\begin{spacing}{0.8}
\begin{algorithm}
\caption{Conditional CDF of the response variable given the explanatory variables with which to predict; based on steps from Algorithms 4, 7, 17, 18 in
Chapter 6 of \citet{joe2014dependence}.}
\label{alg: rvinepcond mix}
\begin{algorithmic}[1]
\State Input: vine array $A = (a_{kj})$ with $a_{jj} = j$ for $j = 1,\ldots,d$ on the diagonal.
    $\mathbf{u}^+=(u_1^+, \ldots, u_d^+)$, $\mathbf{u}^-=(u_1^-, \ldots, u_d^-)$, where
    $u_j^+ = F_j(x_j)$ and $u_j^- = F_j(x_j^-)$ for $1\leq j\leq d-1$, $u_d^+ = u_d^- \in [0,1]$.

\State Output: $\mathbb{P}(F_d(X_d) \leq u_d^+ | X_1 = x_1, \ldots, X_{d-1} = x_{d-1})$.

\State Compute $M = (m_{kj})$ in the upper triangle, where $m_{kj} = \max\{a_{1j},  \ldots  ,a_{kj}\}$ for $k = 1, \ldots , j-1$, $j = 2,\ldots,d$.
\State Compute the $I = (I_{kj} )$ indicator array as in 
Algorithm 5 in \citet{joe2014dependence}.
    
\State $s_{j}^+ = u_{a_{1j}}^+,   s_{j}^- = u_{a_{1j}}^-, w_{j}^+ = u_{j}^+, w_{j}^- = u_{j}^-$, for $j = 1, \ldots, d$.
        
\For {$\ell = 2, \ldots, n_{\mathrm{trunc}}$}
\For {$j = \ell ,\ldots, d$}
\If {$I_{\ell-1,j} = 1$} 
\If {isDiscrete(variable $j$)}
\State  $v_{j}'^+ \leftarrow \frac{C_{a_{\ell-1, j}j; a_{1j} \ldots a_{\ell-2,j} } (s_{j}^+, w_{j}^+ ) - C_{a_{\ell-1, j}j; a_{1j} \ldots a_{\ell-2,j} } (s_{j}^+, w_{j}^- )} {w_{j}^+ - w_{j}^-}$,                
\State  $v_{j}'^- \leftarrow \frac{C_{a_{\ell-1, j}j; a_{1j} \ldots a_{\ell-2,j} } (s_{j}^-, w_{j}^+ ) - C_{a_{\ell-1, j}j; a_{1j} \ldots a_{\ell-2,j} } (s_{j}^-, w_{j}^- )} {w_{j}^+ - w_{j}^-}$,        
\Else
\State  $v_{j}'^+ \leftarrow C_{a_{\ell-1, j} |j; a_{1j} \ldots a_{\ell-2,j} } (s_{j}^+ |w_{j}^+ )$, 
\State $v_{j}'^- \leftarrow C_{a_{\ell-1, j} |j; a_{1j} \ldots a_{\ell-2,j} } (s_{j}^- |w_{j}^+ )$,
\EndIf
\EndIf
\If {isDiscrete(variable $a_{\ell-1,j}$)}
\State  $v_{j}^+ \leftarrow \frac{C_{ a_{\ell-1,j}j; a_{1j} \ldots a_{\ell-2,j}} (s_{j}^+, w_{j}^+) - C_{ a_{\ell-1,j}j; a_{1j} \ldots a_{\ell-2,j}} (s_{j}^-, w_{j}^+)} {s_{j}^+ - s_{j}^-}$,
\State  $v_{j}^- \leftarrow \frac{C_{ a_{\ell-1,j}j; a_{1j} \ldots a_{\ell-2,j}} (s_{j}^+, w_{j}^-) - C_{ a_{\ell-1,j}j; a_{1j} \ldots a_{\ell-2,j}} (s_{j}^-, w_{j}^-)} {s_{j}^+ - s_{j}^-}$,            
\Else
\State  $v_{j}^+ \leftarrow C_{j | a_{\ell-1,j}; a_{1j} \ldots a_{\ell-2,j}} (w_{j}^+ | s_{j}^+ )$,
\State  $v_{j}^- \leftarrow C_{j | a_{\ell-1,j}; a_{1j} \ldots a_{\ell-2,j}} (w_{j}^- | s_{j}^+ )$,
\EndIf
\EndFor
        
\For {$j = \ell+1,\ldots, d$}
\If {$a_{\ell,j} = m_{\ell,j}$}
 $s_{j}^+ \leftarrow v_{m_{\ell+1,j}}^+$,
   $s_{j}^- \leftarrow v_{m_{\ell+1,j}}^-$,
    \ElsIf {$a_{\ell,j} < m_{\ell,j}$}
     $s_{j}^+ \leftarrow v_{m_{\ell+1,j}}'^+$,
       $s_{j}^- \leftarrow v_{m_{\ell+1,j}}'^-$,
\EndIf
\State $w_{j}^+\leftarrow v_{j}^+$,
   $w_{j}^-\leftarrow v_{j}^-$,
\EndFor
\EndFor
\State Return $v_{d}^+$.
\end{algorithmic}
\end{algorithm}
\end{spacing}

\section{Theoretical results on shapes of conditional quantile functions}
\label{sec:theoretical_results}

From the properties of the multivariate normal distribution, if $(X_1, \ldots, X_p, Y)$ follows a multivariate normal distribution, then the conditional quantile function of $Y|X_1, \ldots, X_p$ has the linear form
\[
F_{Y|X_1, \ldots, X_p}^{-1}(\alpha|x_1, \ldots, x_p)= \beta_1 x_1 + \cdots + \beta_p x_p + \Phi^{-1}(\alpha) \sqrt{1 - R^2_{Y; X_1, \ldots, X_p}},
  \quad 0<\alpha<1, 
\]
where $R^2_{Y; X_1, \ldots, X_p}$ is the multiple correlation coefficient.
Going beyond the normal distribution, we address the following question in this section: how does the choice of bivariate copulas in a vine copula regression model affect the conditional quantile function, especially when the explanatory variables are large (in absolute value)?
For comparisons with multivariate normal, we assume the variables
$X_1,\ldots,X_p,Y$ have been transformed so that they have marginal
$N(0,1)$ distributions.
In this case, plots from vine copulas with one or two explanatory variables can
show conditional quantile functions that are close to linear in the middle,
and asymptotically linear, sublinear or constant along different directions
to $\pm\infty$; \citet{bernard2015conditional} have several figures show
this pattern for the case of one explanatory variable.
Such behavior cannot be
obtained with regression equations that are linear in $\beta$'s and
is hard to obtain with
nonlinear regression functions that are directly specified.

We start with the bivariate case (one explanatory variable). 
Conditions are obtained to classify the asymptotic behavior of conditional quantile function into four categories: \textit{strongly linear, weakly linear, sublinear} and \textit{asymptotic constant}.
For bivariate Archimedean copulas, the conditions are related to
conditions on the Laplace transform generator.
In the supplementary material, we have further results for the trivariate case $F_{Y|X_1, X_2}^{-1}(\alpha | x_1, x_2)$ with a trivariate vine copula and briefly discuss the possibility of generalizing the results to higher dimensions.

We focus on a bivariate random vector $(X, Y)$ with standard normal margins. Let $C(u, v)$ be the copula, then the joint CDF is $F_{X,Y}(x, y) = C(\Phi(x), \Phi(y))$. 
Without loss of generality, we assume the copula $C(u, v)$ has positive dependence.
We are interested in the shape of the conditional CDF $F_{Y|X}(y|x)$ and conditional quantile $F_{Y|X}^{-1}(\alpha|x)$, when $x$ is extremely large or small and $\alpha\in(0,1)$ is fixed.
\citet{bernard2015conditional} study a few special cases for bivariate copulas.
Our results are more extensive in relating the shape of asymptotic quantiles to the strength of dependence in the joint tail.

If the conditional distribution $C_{V|U}(\cdot|u)$ converges to a continuous
distribution with support on $[0,1]$, as $u \to 0^+$, then
$C_{V|U}^{-1}(\alpha|0) > 0$ , for $\alpha \in (0,1)$. Therefore,
$F_{Y|X}^{-1}(\alpha|x)$ levels off as $x \to -\infty$. 
The same argument applies when $x \to +\infty$. 
That is,
\begin{align*}
\lim_{x\to -\infty} F_{Y|X}^{-1}(\alpha|x) &= \Phi^{-1}(C_{V|U}^{-1}(\alpha|0)),
\\
\lim_{x\to +\infty} F_{Y|X}^{-1}(\alpha|x) &= \Phi^{-1}(C_{V|U}^{-1}(\alpha|1)).
\end{align*}

If $\lim_{u\to0^+} C_{V|U}(\cdot|u)$ is degenerate at 0,
then $\lim_{u \to 0^+} C_{V|U}^{-1}(\alpha|u) = 0$. To study the shape of $F_{Y|X}(y|x)$ when $x$ is 
very negative,
we need to further investigate the rate at which $C_{V|U}^{-1}(\alpha|u)$ converges to 0.
The next proposition, with proof in the supplementary material,
summarizes the possibilities.

\begin{proposition}
\label{prop: asympt cond}
Let $(X, Y)$ be a bivariate random vector with standard normal margins and a positively dependent copula $C(u, v)$.
\begin{itemize}

\item (Lower tail)     Fixing $\alpha \in (0, 1)$, if $-\log C_{V|U}^{-1}
(\alpha | u) \sim k_{\alpha} (-\log u)^{\eta}$ as $u \to 0^+$, then
$F_{Y|X}^{-1}(\alpha|x) \sim -(2^{1-\eta} k_{\alpha})^{1/2} |x|^{\eta}$ as
$x \to -\infty$.

\item (Upper tail)     Fixing $\alpha \in (0, 1)$, if $-\log[1 -
C_{V|U}^{-1} (\alpha | u)] \sim k_{\alpha} [-\log (1-u)]^{\eta}$ as $u \to
1^-$, then $F_{Y|X}^{-1}(\alpha|x) \sim (2^{1-\eta} k_{\alpha})^{1/2}
x^{\eta}$ as $x \to +\infty$.

\end{itemize}
\end{proposition}

\begin{figure}
\centering
\includegraphics[width=0.6\textwidth]{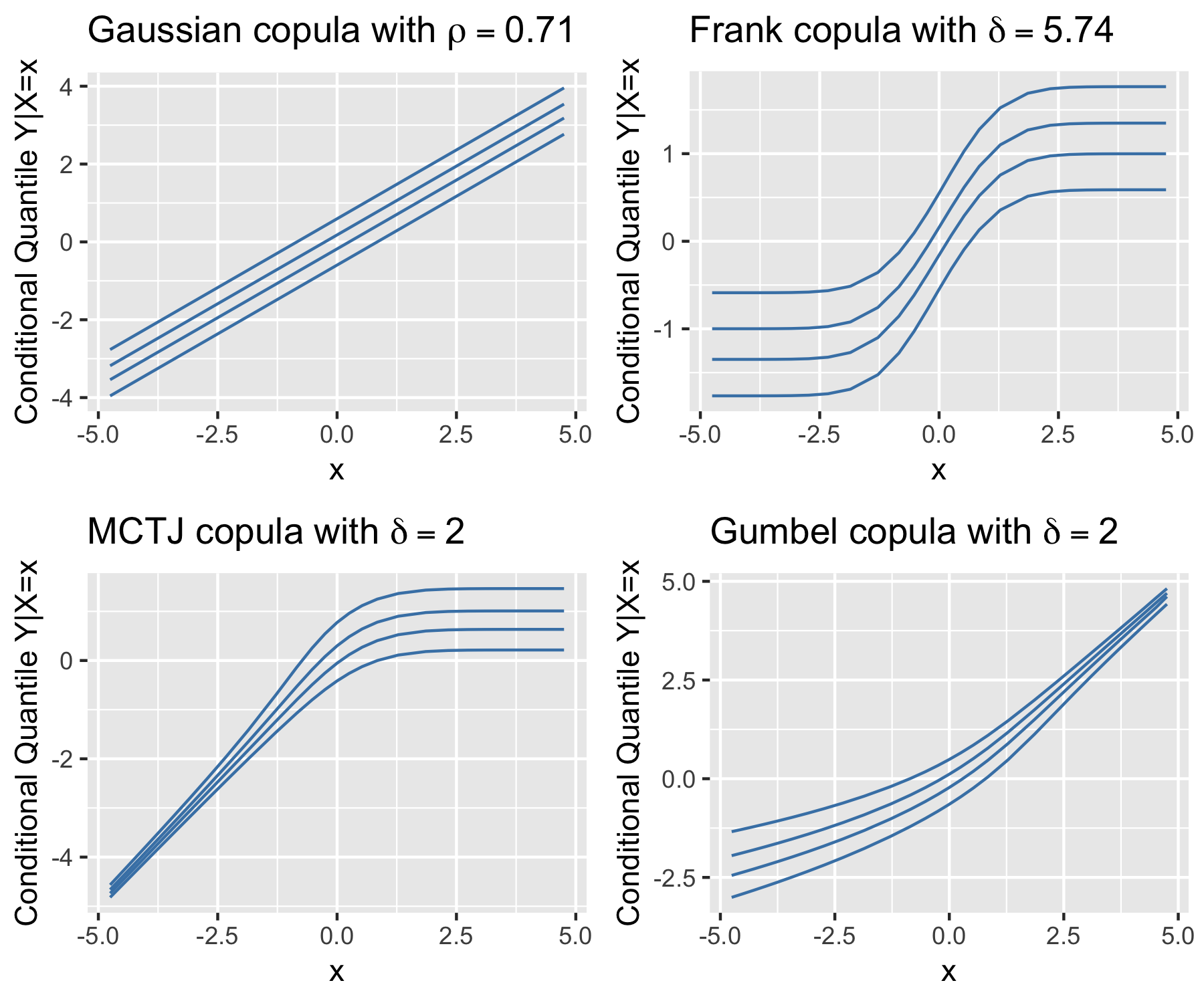}
\caption{Conditional quantile functions for bivariate copulas with Kendall's
$\tau=0.5$, combined with $N(0,1)$ margins. The quantile levels are $20\%, 40\%, 60\%$ and $80\%$. 
The parameter of the corresponding copula is denoted by $\rho$ or $\delta$.
This figure is adapted from \citet{bernard2015conditional}.
}
\label{fig:bivariate_cond_quan}   
\end{figure}

Here $\eta$ indicates the strength of relation between two variables in the tail; a larger $\eta$ value corresponds to stronger relation.
The strongest possible comonotonic dependence is when $Y=X$, and the conditional quantile function is $F_{Y|X}^{-1}(\alpha|x)=x$, which is linear in $x$ and does not depend on $\alpha$; in this case, $\eta=1$.
The weakest possible positive dependence is when $X$ and $Y$ are independent, and $F_{Y|X}^{-1}(\alpha|x)=F_{Y}^{-1}(\alpha)$ does not depend on $x$; in this case, $\eta=0$.
Based on the value of $\eta$, the asymptotic behavior of the conditional quantile function can be classified into the following categories: 
\begin{enumerate}

    \item Strongly linear: $\eta=1$ and $k_{\alpha} = 1$. $F_{Y|X}^{-1}(\alpha|x)$ goes to infinity linearly, and it does not depend on $\alpha$. It has stronger dependence than bivariate normal.
    
    \item Weakly linear: $\eta=1$, $ k_{\alpha}$ can depend on $\alpha$
    and $0<k_\alpha<1$. $F_{Y|X}^{-1}(\alpha|x)$ goes to infinity linearly and it depends on $\alpha$. It has comparable dependence with bivariate normal.
    
    \item Sublinear: $0<\eta<1$. $F_{Y|X}^{-1}(\alpha|x)$ goes to infinity sublinearly.
    The dependence is weaker than bivariate normal.
    
    \item Asymptotically constant: $\eta=0$. $F_{Y|X}^{-1}(\alpha|x)$ converges to a finite constant. Asymptotically it behaves like independent.
\end{enumerate}
\autoref{fig:bivariate_cond_quan} shows the conditional quantile functions for bivariate copulas with different $\eta$ in the upper and lower tails.
Example~\ref{eg: MTCJ lower quantile} to~\ref{eg: Gumbel lower tail} derive the conditional quantile functions for bivariate Mardia-Takahasi-Clayton-Cook-Johnson (MTCJ) and Gumbel copulas.
Note that $\eta$ is constant over $\alpha$ for several commonly used
parametric bivariate copula families. 
However, there are cases where $\eta$ depends on $\alpha$.
For example, the boundary conditional distribution of the bivariate Student-$t$ copula has mass at both 0 and 1; depending on the value of $\alpha$, $C^{-1}_{V|U}(\alpha|u)$ could go to either 0 or 1, as $u \to 0$.

\begin{example}
\label{eg: MTCJ lower quantile}
(MTCJ lower tail)
The bivariate MTCJ copula CDF is
\[
    C(u, v; \delta) = (u^{-\delta} + v^{-\delta} - 1)^{-1/\delta},
    \quad 0<u<1, 0<v<1, \delta>0.
\]
The conditional quantile function is
\begin{equation*}
C_{V|U}^{-1}(\alpha | u; \delta) 
= 
[(\alpha^{-\delta/(1+\delta)}-1)u^{-\delta} + 1]^{-1/\delta}
\sim
(\alpha^{-\delta/(1+\delta)}-1)^{-1/\delta}u,
\quad u \to 0.
\end{equation*}
Take the log of both sides,
$-\log C_{V|U}^{-1}(\alpha | u; \delta) 
\sim
\log u$.
By Proposition~\ref{prop: asympt cond}, we have $F_{Y|X}^{-1}(\alpha|x) 
\sim
x$,
as $x \to -\infty$. 
To apply the next proposition to get the
same conclusion, the generator is the gamma Laplace transform
$\psi(s)=(1+s)^{-1/\delta}$.
\end{example}

\begin{example} 
\label{eg: Gumbel lower tail}
(Gumbel lower tail)
The bivariate Gumbel copula CDF is
\[
    C(u, v; \delta) = \exp\{ - [ (-\log u)^\delta + (-\log v)^\delta ]^{1/\delta} \},
    \quad 0<u<1, 0<v<1, \delta>1.
\]
The conditional CDF is 
\[
C_{V|U}(v|u;\delta) = u^{-1} \exp \bigl\{-[(-\log u)^{\delta} + (-\log v)^{\delta}]^{1/\delta}\bigr\} \Bigl[1+\Bigl(\frac{-\log v}{-\log u}\Bigr)^{\delta}\Bigr]^{1/\delta-1}.
\]
The conditional quantile function $C_{V|U}^{-1}(\alpha|u;\delta)$ does not have a closed-form expression;
it has the following asymptotic expansion:
\begin{equation*}
- \log C_{V|U}^{-1}(\alpha|u; \delta)
\sim (-\delta \log \alpha)^{1/\delta}(-\log u)^{1-1/\delta},
\quad u\to 0.
\end{equation*}
By Proposition~\ref{prop: asympt cond}, we have 
$F_{Y|X}^{-1}(\alpha|x) 
\sim -(-2\delta \log \alpha)^{1/(2\delta)}\, |x|^{1-1/\delta}$, 
as $x\to -\infty$.
To apply the next proposition to get the
same conclusion, the generator is the positive stable Laplace transform
$\psi(s)=\exp\{-s^{-1/\delta}\}$.
\end{example}

For Archimedean and survival Archimedean copulas, the following proposition provides some links between tail dependence behavior and tail conditional distribution and quantile functions.
The proof of the proposition is included in
the supplementary material.

\begin{proposition}
\label{prop: Archimedean tail functions}
Given the generator or Laplace transform (LT) $\psi$ of an Archimedean copula, 
we assume the following.
\begin{enumerate}

\item For the upper tail of $\psi$,
as $s \to +\infty$, 
\begin{equation}
    \label{eq:psi_expansion_lower_maintext}
    \psi(s) \sim T(s) = a_1 s^q \exp(-a_2 s^r) \quad \mathrm{and} \quad \psi'(s) \sim T'(s), 
\end{equation}
where $a_1 > 0$, $r=0$ implies $a_2 = 0$ and $q<0$, and $r > 0$ implies $r \leq 1$ and $q$ can be 0, negative or positive.    

\item For the lower tail of $\psi$, as $s\to 0^+$, there is $M\in(k,k+1)$ such that
\begin{equation}
    \label{eq:psi_expansion_upper_maintext}
    \psi(s) = \sum_{i=0}^{k} (-1)^i h_i s^i + (-1)^{k+1} h_{k+1} s^M +
o(s^M), \quad s\to 0^+,
\end{equation}
where $h_0=1$ and  $0 < h_i < \infty$ for $i = 1, \ldots, k+1$.
If $0<M<1$, then $k=0$.
\end{enumerate}

Then we have the following.

\begin{itemize}

\item (Lower tail)     
If $v\in(0,1)$ and $\alpha \in (0,1)$ are fixed, then as $u\to 0$,  
\begin{equation}
    C_{V|U}(v|u) \sim 
  \begin{cases}
    1 + (q-1) \psi^{-1}(v) \left( u / a_1 \right)^{-1/q} \to 1       & \quad \text{if } r=0,\\
    1 - a_2^{1/r} r \psi^{-1}(v) (-\log u)^{1 - 1/r} \to 1      & \quad \text{if } 0 < r < 1,\\
    \mathrm{const} \in (0, 1)       & \quad \text{if } r=1.\\
  \end{cases}
  \label{eq:prop_4_2_lower_cdf}  
\end{equation}

\begin{equation}
    C_{V|U}^{-1}(\alpha|u) \sim 
  \begin{cases}
    \left( \alpha^{1/(q-1)} - 1 \right)^q \cdot u \to 0       & \quad \text{if } r=0,\\
    \exp \left[ -\left(-r^{-1} \log\alpha\right)^r (-\log u)^{1-r} \right] 
    \to 0      & \quad \text{if } 0 < r < 1,\\
    \mathrm{const} \in (0, 1)       & \quad \text{if } r=1.\\
  \end{cases}
  \label{eq:prop_4_2_lower_quantile}
\end{equation}
The cases $r=0$, $0<r<1$ and $r=1$ correspond to lower tail 
dependence, intermediate dependence and quadrant independence respectively.

\item (Upper tail)    If  $v\in(0,1)$ and $\alpha \in (0,1)$ are fixed, then 
as $u\to 1$, 
\begin{equation}
    C_{V|U}(v|u) \sim 
  \begin{cases}
    -\frac{\psi'(\psi^{-1}(v))}{h_1^{1/M} M}
    (1-u)^{(1-M)/M}
    \to 0       & \quad \text{if } 0 < M < 1,\\
    \mathrm{const} \in (0, 1)       & \quad \text{if } M>1.\\
  \end{cases}
  \label{eq:prop_4_2_upper_cdf}    
\end{equation}

\begin{equation}
    C_{V|U}^{-1}(\alpha|u) \sim 
  \begin{cases}
    1 - \left( \alpha^{1/(M-1)} - 1 \right)^M (1-u)
    \to 1       & \quad \text{if } 0 < M < 1,\\
    \mathrm{const} \in (0, 1)       & \quad \text{if } M>1.\\
  \end{cases}
  \label{eq:prop_4_2_upper_quantile}
\end{equation}
The cases $0<M<1$ and $M>1$ correspond to upper tail dependence and 
intermediate dependence/quadrant independence, respectively.
Note that we do not cover the case of $M=1$ for the upper tail because it involves a slowly varying function. 

\end{itemize}
\end{proposition}

Combined with Proposition~\ref{prop: asympt cond}, the above proposition states that, for the lower tail, the three cases $r=0$, $0<r<1$ and $r=1$ correspond to strongly linear, sublinear and asymptotic constant conditional quantile functions respectively; for the upper tail, the two cases $0<M<1$ and $M>1$ correspond to strongly linear and asymptotic constant conditional quantile functions respectively.

For trivariate vine copula models, the asymptotic behavior of conditional quantile 
functions also have the four shapes: \textit{strongly linear}, 
\textit{weakly linear}, \textit{sublinear} and \textit{asymptotically constant}.
However, extending from bivariate to trivariate is not trivial, since the asymptotic conditional quantile function depends on the direction in which the covariates go to infinity.
The trivariate case provides insight on the type of asymptotic behavior in higher dimensions. 
See the supplementary material for a detailed analysis.

\section{Simulation study}
\label{sec: Simulation_Study}

We demonstrate the flexibility and effectiveness of vine copula regression methods by visualizing the fitted models on simulated datasets.
The simulated datasets have three variables: $X_1$ and $X_2$ are the
explanatory variables and $Y$ is the response variable, where 
\[
\mathbf{X} = \begin{pmatrix} X_1 \\ X_2 \end{pmatrix}
\sim N \left( 
\begin{pmatrix} 0 \\ 0 \end{pmatrix},
\begin{pmatrix} 1 & 0.5 \\ 0.5 & 1 \end{pmatrix}
\right)
\]
and $Y$ is simulated in three cases with varying conditional expectation and variance structures. Let $U_1 = \Phi(X_1)$ and $U_2 = \Phi(X_2)$, where $\Phi$ is the standard normal CDF,
and $\epsilon$ be a random error following a standard normal distribution and independent from $X_1$ and $X_2$. 
The three cases are as follows:
\begin{enumerate}
  \item Linear and homoscedastic: $Y = 10 X_1 + 5 X_2 + 10 \epsilon$.
  \item Linear and heteroscedastic: $Y = 10 X_1 + 5 X_2 + 10 (U_1 + U_2) \epsilon$.
  \item Non-linear and heteroscedastic: $Y = U_1 e^{1.8 U_2} + 0.5 (U_1 + U_2) \epsilon$.
\end{enumerate}

We simulate samples with size 2000 in each case with a random split
of 1000 observations for a training set and a test set.
Five methods are considered in the simulation study: (1) linear
regression, (2) linear regression with logarithmic transformation of the 
response variable, (3) quadratic regression, (4) Gaussian copula regression, and 
(5) vine copula regression. 
The Gaussian copula can be considered as a special case of the vine copula, 
in which the bivariate copula families on the vine edges are
all bivariate Gaussian.
Different models are trained on the training set and used to obtain the
conditional expectations as point predictions and 95\% prediction
intervals on the test set.
For copula regressions, the upper and lower bounds of the 95\% prediction interval are the conditional 97.5\% and 2.5\% quantiles respectively.
For the Gaussian and vine copula, the marginal distribution of
$Y$ is fitted by the MLE of a normal distribution in case 1. 
In cases 2 and 3, the distributions of the response variable are
skewed and unimodal but not too heavy-tailed.
Therefore, we fit 3-parameter skew-normal distributions.
For the vine copula regression, the candidate bivariate copula families include 
Student-$t$, MTCJ, Gumbel, Frank, Joe, BB1, BB6, BB7, BB8, and the corresponding survival copulas.
The bivariate copulas are selected using the AIC described in 
\autoref{sec: Bivariate_Copula_Selection}.
The procedure is replicated 100 times and the average scores
of the replicates are reported in \autoref{tab: simulation_comparison_dim3}.
To evaluate the performance of a regression model, we apply the
root-mean-square error (RMSE) and several scoring rules for probabilistic 
forecasts studied in \citet{gneiting2007strictly}, including the logarithmic
score (LogS), quadratic score (QS), interval score (IS), and integrated Brier score (IBS).
Note that the RMSE is not meaningful if there is heteroscedasticity in 
conditional distributions; 
the LogS, QS, IS, and IBS assess the predictive distributions with non-constant 
variance more effectively.

\begin{itemize}
  \item 
    The root-mean-square error (RMSE) measures a model's performance on point estimations.
    \[
    \operatorname{RMSE}(\mathcal{M}) = \sqrt{\frac{1}{n_{\mathrm{test}}}\sum_{i=1}^{n_{\mathrm{test}}} (y_i - \hat{y}_i^{\mathcal{M}})^2},
    \]
    where $y_i$ is the response variable of the $i$-th sample in the test set, and $\hat{y}_i^{\mathcal{M}}$is the predictive conditional expectation of a fitted model $\mathcal{M}$.
  \item 
    The logarithmic score (LogS) is a scoring rule for probabilistic forecasts of continuous variables \citep{gneiting2007strictly}.
    It is closely related to the generalization error in machine learning literature (Chapter 7.2 in~\citet{hastie09elements}).    
    \[
        \operatorname{LogS}(\mathcal{M}) = 
        \frac{1}{n_{\mathrm{test}}}
        \sum_{i=1}^{n_{\mathrm{test}}} 
        \log \hat{f}_{Y|\mathbf{X}}^{\mathcal{M}} (y_i | \mathbf{x}_i),
    \]
    where $(\mathbf{x}_i, y_i)$ is the $i$th observation in the test set, and $\hat{f}_{Y|\mathbf{X}}^{\mathcal{M}}$ is the predictive conditional PDF of model $\mathcal{M}$. 
    For example, if $\mathcal{M}$ is a linear regression, then the predictive conditional distribution is a scaled and shifted $t$-distribution.
    If $\mathcal{M}$ is a vine copula, the predictive conditional distribution can be calculated using the procedure described in \autoref{sec:prediction}.

  \item
    The quadratic score (QS) measures the predictive density, penalized by 
    its $\mathcal{L}_2$ norm \citep{gneiting2007strictly}:
        \[
        \operatorname{QS}(\mathcal{M}) = 
        \frac{1}{n_{\mathrm{test}}}
        \sum_{i=1}^{n_{\mathrm{test}}}
        \left[ 
        2\hat{f}_{Y|\mathbf{X}}^{\mathcal{M}} (y_i | \mathbf{x}_i)
        -
        \int_{-\infty}^{\infty} 
        \hat{f}_{Y|\mathbf{X}}^{\mathcal{M}} (y | \mathbf{x}_i)^2 \, \mathrm{d}y
        \right].
    \]
  
  \item
        The interval score (IS) is a scoring rule for quantile and interval 
        forecasts \citep{gneiting2007strictly}. 
        In the case of the central $(1-\alpha) \times 100\%$ prediction 
        interval, let $\hat{u}_i^{\mathcal{M}}$ and $\hat{\ell}_i^{\mathcal{M}}$
        be the predictive quantiles at level $\alpha / 2$ and $1-\alpha / 2$ by
        model $\mathcal{M}$ for the $i$-th test sample.
        The interval score of model $\mathcal{M}$ is 
        \begin{multline*}
        \operatorname{IS}(\mathcal{M}) = 
        \frac{1}{n_{\mathrm{test}}}
        \sum_{i=1}^{n_{\mathrm{test}}}
        \Big[
        (\hat{u}_i^{\mathcal{M}} - \hat{\ell}_i^{\mathcal{M}})
        \\
        +
        \frac{2}{\alpha} (\hat{\ell}_i^{\mathcal{M}} - y_i) 
        \mathbb{I}\{y_i < \hat{\ell}_i^{\mathcal{M}}\}
        +
        \frac{2}{\alpha} (y_i - \hat{u}_i^{\mathcal{M}}) 
        \mathbb{I}\{y_i > \hat{u}_i^{\mathcal{M}}\}        
        \Big].
        \end{multline*}
        Smaller interval scores are better. 
        A model is rewarded for narrow prediction intervals, and it incurs a penalty, the size of which depends on $\alpha$, if an observation 
        misses the interval.
  
  \item    
    The integrated Brier score (IBS) is a scoring rule that is defined in terms of predictive cumulative distribution functions \citep{gneiting2007strictly}:
    \[
        \operatorname{IBS}(\mathcal{M}) = 
        \frac{1}{n_{\mathrm{test}}}
        \sum_{i=1}^{n_{\mathrm{test}}}
        \int_{-\infty}^{\infty} 
        \left[
        \widehat{F}_{Y|\mathbf{X}}^{\mathcal{M}} (y | \mathbf{x}_i) 
        -       
        \mathbb{I}\{y \geq y_i\}
        \right]^2
        \, \mathrm{d}y,
    \]    
    where $\widehat{F}_{Y|\mathbf{X}}^{\mathcal{M}}$ is the predictive conditional CDF of model $\mathcal{M}$.
    Smaller integrated Brier scores are better.
\end{itemize}

\begin{figure}
\centering
\begin{subfigure}[t]{0.4\textwidth}
    \centering
    \includegraphics[width=\textwidth]{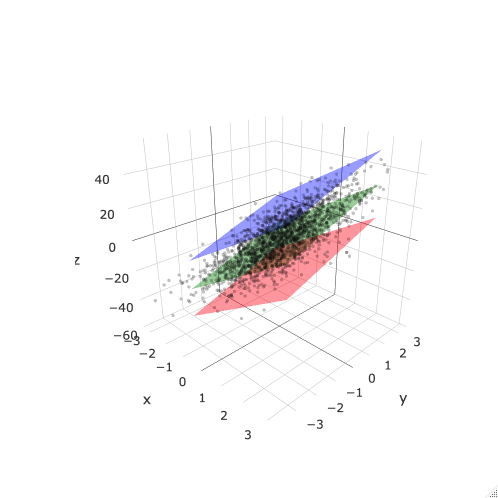}
    \caption{Linear and homoscedastic data, the true surfaces.}
    \label{fig:linear_homo_true}   
\end{subfigure}%
\qquad
\begin{subfigure}[t]{0.4\textwidth}
    \centering
    \includegraphics[width=\textwidth]{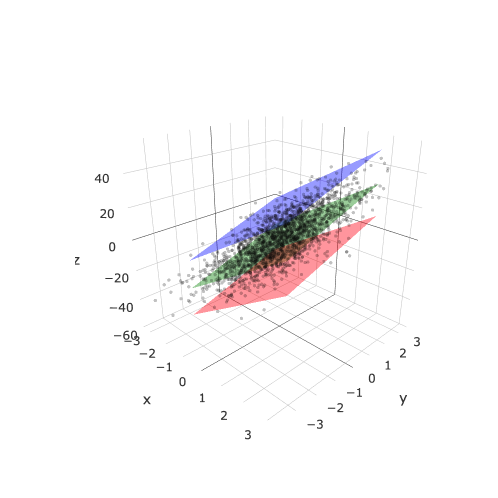}
    \caption{Linear and homoscedastic data, predicted surfaces by a vine copula regression model. 
    }
    \label{fig:linear_homo_pred}   
\end{subfigure}
\caption{The linear homoscedastic simulation case. 
In this fitted vine copula model, $C_{13}, C_{12}$ and $C_{23;1}$ are all Gaussian copulas, with parameters $\rho_{13} = 0.77, \rho_{12} = 0.5$ and $\rho_{23;1} = 0.39$.
}
\end{figure}

The first case serves as a sanity check; if the response variable
is linear in the explanatory variables and the conditional variance is
constant, the vine copula should behave like linear regression.
\autoref{fig:linear_homo_true} plots the simulated data, the
true conditional expectation surface and true 95\% prediction interval surfaces. 
\autoref{fig:linear_homo_pred} plots the corresponding predicted surfaces.
All three surfaces truthfully reflect the linearity of the data.
The first three lines of \autoref{tab: simulation_comparison_dim3} show 
that the vine copula and linear regression have similar performance 
in terms of all five metrics.

\begin{figure}
\centering
\begin{subfigure}[t]{0.4\textwidth}
    \centering
    \includegraphics[width=\textwidth]{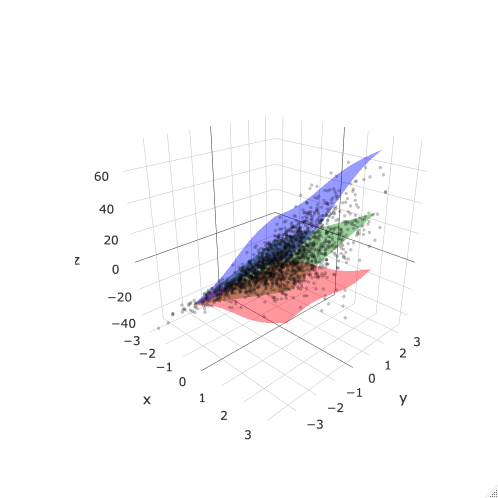}
    \caption{Linear and heteroscedastic data, the true surfaces.}
    \label{fig:linear_hetero_true}   
\end{subfigure}%
\qquad
\begin{subfigure}[t]{0.4\textwidth}
    \centering
    \includegraphics[width=\textwidth]{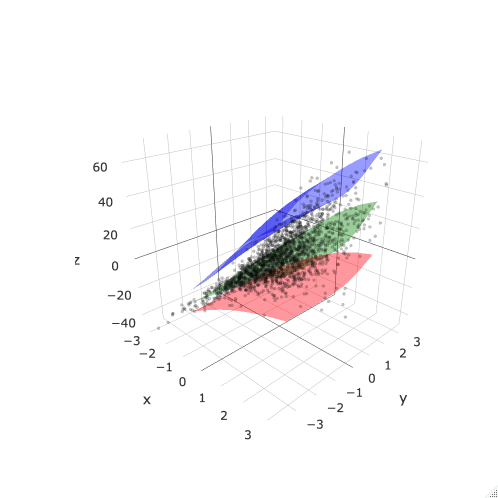}
    \caption{Linear and heteroscedastic data, predicted surfaces by a vine copula regression model.}
    \label{fig:linear_hetero_pred}   
\end{subfigure}
\caption{The linear heteroscedastic simulation case.
In this fitted vine copula model, 
$C_{13}$ is a survival Gumbel copula with parameter $\delta_{13} = 2.21$, 
$C_{12}$ is a Gaussian copula with parameter $\rho_{12} = 0.5$,
and $C_{23;1}$ is a BB8 copula with parameters $\vartheta_{23;1} = 3.06, \delta_{23;1} = 0.71$.
}
\end{figure}

The second case adds heteroscedasticity to the first case; that is,
the variance of $Y$ increases as $X_1$ or $X_2$ increases while the
linear relationship remains the same. We expect the conditional expectation
surface to be linear. 
\autoref{fig:linear_hetero_true} and \autoref{fig:linear_hetero_pred} show the true and predicted surfaces respectively.
The conditional expectation surface is linear and the lengths of prediction intervals increase with $X_1$ and $X_2$.
The performance measures in \autoref{tab: simulation_comparison_dim3} are also
consistent with our expectation: the vine copula models have better LogS, QS, IS, and IBS, although the RMSE is slightly worse than the linear regression model. 
The logarithmic transformation of the response variable does not seem to
improve the performance.

\begin{figure}
\centering
\begin{subfigure}[t]{0.4\textwidth}
  \centering
  \includegraphics[width=\textwidth]{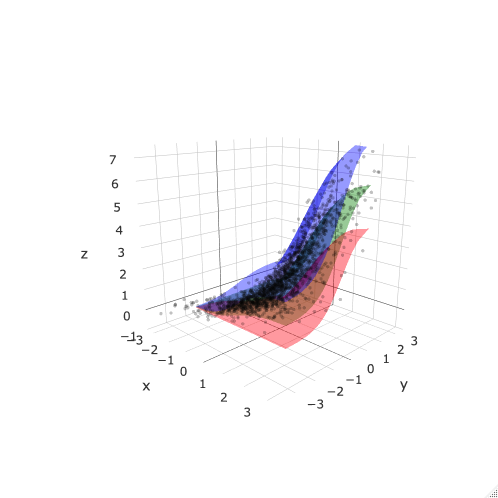}
  \caption{Non-linear and heteroscedastic data, the true surfaces.}
  \label{fig:nl_hetero_true}
\end{subfigure}%
\qquad
\begin{subfigure}[t]{0.4\textwidth}
  \centering
  \includegraphics[width=\textwidth]{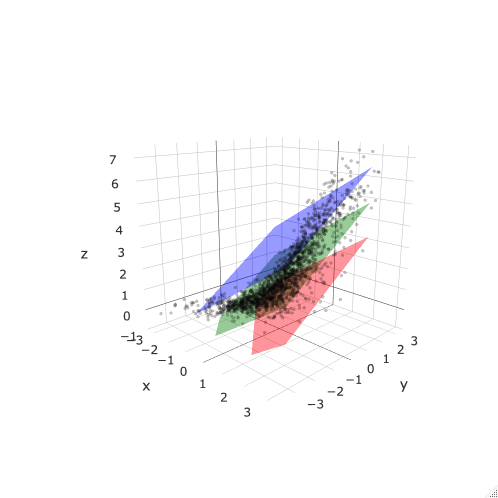}
  \caption{Non-linear and heteroscedastic data, predicted surfaces by a linear regression model.}
  \label{fig:nl_hetero_pred_lm}
\end{subfigure}
\\
\begin{subfigure}[t]{0.4\textwidth}
  \centering
  \includegraphics[width=\textwidth]{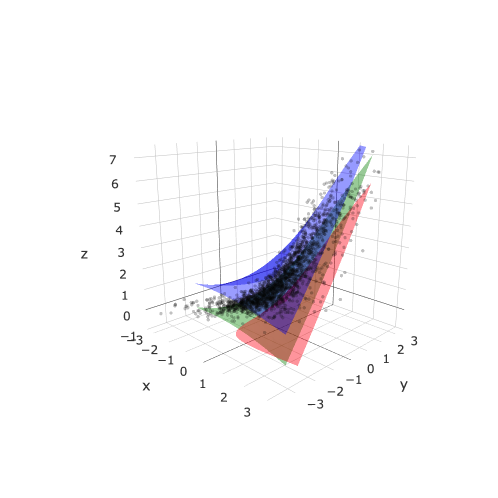}
  \caption{Non-linear and heteroscedastic data, predicted surfaces by a quadratic regression model.}
  \label{fig:nl_hetero_pred_quad}
\end{subfigure}  
\qquad
\begin{subfigure}[t]{0.4\textwidth}
  \centering
  \includegraphics[width=\textwidth]{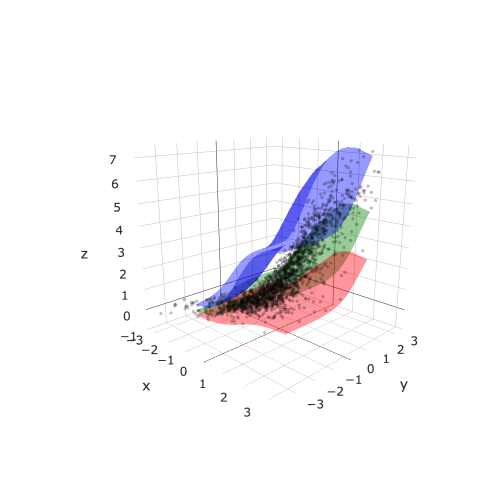}
  \caption{Non-linear and heteroscedastic data, predicted surfaces by a vine copula regression model.}
  \label{fig:nl_hetero_pred_vine}
\end{subfigure}
\caption{The non-linear and heteroscedastic simulation case.
In this fitted vine copula model, 
$C_{13}$ is a survival BB8 copula with parameters $\vartheta_{13} = 6, \delta_{13} = 0.78$, 
$C_{12}$ is a Gaussian copula with parameter $\rho_{12} = 0.5$,
and $C_{23;1}$ is a BB8 copula with parameters $\vartheta_{23;1} = 6, \delta_{23;1} = 0.65$.
}
\label{fig:sim_nonlinear_hetero}
\end{figure}

Finally, the third case incorporates both non-linearity and heteroscedasticity. 
Since the linear regression obviously cannot fit the non-linear trend, we compare our model to quadratic regression as well.
\autoref{fig:sim_nonlinear_hetero} shows the true surfaces and the predicted surfaces for the three models.
Although the quadratic regression model captures the
non-linear trend, it is not flexible enough to model
heteroscedasticity. Another drawback of quadratic regression is that, the
conditional mean $\hat{y}$ is not always monotonically increasing with
respect to $x_1$ and $x_2$, and this contradicts the pattern in the data.
The vine copula naturally fits the non-linearity and heteroscedasticity pattern.
Quantitatively, 
the quadratic regression model has the best RMSE and IS, but vine copula
models have the best LogS, QS, and IBS, as shown in \autoref{tab: simulation_comparison_dim3}.

\begin{table}
\tabcolsep=4pt 
\footnotesize
\centering
\begin{tabular}{llrrrrr}
\toprule
Case                 & Model     & RMSE$\,\downarrow$  & LogS$\,\uparrow$ & QS$\,\uparrow$ & IS$\,\downarrow$ & IBS$\,\downarrow$ \\ 
\midrule
\multirow{2}{*}{1}                 & Linear reg.          & 10.01 (0.02) & $-3.72$ (0.00) & 0.028 (0.000) & 39.25 (0.09) & 5.64 (0.01)   \\ \cmidrule{2-7}
                                   & Gaussian copula reg. & 10.01 (0.02) & $-3.72$ (0.00) & 0.028 (0.000) & 39.09 (0.09) & 5.64 (0.01)   \\ \cmidrule{2-7}
                                   & Vine copula reg.     & 10.01 (0.02) & $-3.72$ (0.00) & 0.028 (0.000) & 39.14 (0.09) & 5.64 (0.01)   \\ \midrule
\multirow{2}{*}{2}                 & Linear reg.          & 11.19 (0.03) & $-3.83$ (0.00) & 0.028 (0.000) & 43.80 (0.14) & 6.06 (0.02)   \\ \cmidrule{2-7}
                                   & Reg.~with log-transform   & 11.71 (0.04) & $-3.83$ (0.01) & 0.031 (0.000) & 47.22 (0.30) & 6.09 (0.02)   \\ \cmidrule{2-7}
                                   & Gaussian copula reg. & 11.32 (0.03) & $-3.75$ (0.00) & 0.031 (0.000) & 41.45 (0.13) & 5.95 (0.02)   \\ \cmidrule{2-7}
                                   & Vine copula reg.     & 11.38 (0.03) & $-3.73$ (0.00) & 0.033 (0.000) & 41.24 (0.12) & 5.97 (0.02)  \\ \midrule
\multirow{3}{*}{3}                 & Linear reg.          & 0.77 (0.01)  & $-1.16$ (0.00) & 0.388 (0.001) & 3.03 (0.00) & 0.43 (0.00)   \\ \cmidrule{2-7}
                                   & Reg.~with log-transform   & 0.69 (0.00) & $-0.87$ (0.00) & 0.540 (0.002) & 2.52 (0.01) & 0.35 (0.00)   \\ \cmidrule{2-7}
                                   & Quadratic reg.       & 0.62 (0.00)  & $-0.95$ (0.00) & 0.511 (0.001) & 2.43 (0.01) & 0.34 (0.00)   \\ \cmidrule{2-7}
                                   & Gaussian copula reg. & 0.69 (0.00)  & $-0.86$ (0.00) & 0.604 (0.002) & 2.65 (0.01) & 0.35 (0.00)   \\ \cmidrule{2-7}
                                   & Vine copula reg.     & 0.63 (0.00)  & $-0.75$ (0.00) & 0.686 (0.002) & 2.50 (0.01) & 0.32 (0.00)   \\ \bottomrule
\end{tabular}
\caption{
Simulation results for two explanatory variables.
The table shows the root-mean-square error (RMSE), logarithmic score (LogS), 
quadratic score (QS), interval score (IS), and integrated Brier score (IBS) in different simulation cases. 
The arrows in the header indicate that lower RMSE, IS, and IBS; 
and higher LogS and QS are better.
The numbers in parentheses are the corresponding standard errors.}
\label{tab: simulation_comparison_dim3}
\end{table}

We have also conducted a similar simulation study with four explanatory variables
$X_1, X_2, X_3, X_4$, where
\[
\mathbf{X} = \begin{pmatrix} X_1 \\ X_2 \\ X_3 \\ X_4 \end{pmatrix}
\sim N \left( 
\begin{pmatrix} 0 \\ 0 \\ 0 \\ 0 \end{pmatrix},
\begin{pmatrix} 
1 & 0.5 & 0.5 & 0.5 \\ 
0.5 & 1 & 0.5 & 0.5 \\ 
0.5 & 0.5 & 1 & 0.5 \\
0.5 & 0.5 & 0.5 & 1
\end{pmatrix}
\right).
\]
The response variable $Y$ is generated from similar three cases:
\begin{enumerate}
  \item Linear and homoscedastic: 
  $Y = 5 (X_1 + X_2 + X_3 + X_4) + 20 \epsilon$.
  \item Linear and heteroscedastic: 
  $Y = 5 (X_1 + X_2 + X_3 + X_4) + 10 (U_1 + U_2 + U_3 + U_4) \epsilon$.
  \item Non-linear and heteroscedastic: 
  $Y = U_1 U_2 e^{1.8 U_3 U_4} + 0.5 (U_1 + U_2 + U_3 + U_4) \epsilon$.
\end{enumerate}
The results of the simulation study are shown in \autoref{tab: simulation_comparison_dim5},
the pattern of which is similar to that of \autoref{tab: simulation_comparison_dim3}.

\begin{table}
\tabcolsep=4pt 
\footnotesize
\centering
\begin{tabular}{llrrrrr}
\toprule
Case                 & Model     & RMSE$\,\downarrow$  & LogS$\,\uparrow$ & QS$\,\uparrow$ & IS$\,\downarrow$ & IBS$\,\downarrow$ \\ 
\midrule
\multirow{2}{*}{1}                 & Linear reg.          & 20.09 (0.05) & $-4.42$ (0.00) & 0.014 (0.000) & 78.53 (0.18) & 11.34 (0.03)   \\ \cmidrule{2-7}
                                   & Gaussian copula reg. & 20.09 (0.05) & $-4.42$ (0.00) & 0.014 (0.000) & 78.06 (0.18) & 11.34 (0.03)  \\ \cmidrule{2-7}
                                   & Vine copula reg.     & 20.12 (0.05) & $-4.42$ (0.00) & 0.014 (0.000) & 78.18 (0.18) & 11.36 (0.03)   \\ \midrule
\multirow{2}{*}{2}                 & Linear reg.          & 22.04 (0.07) & $-4.51$ (0.00) & 0.014 (0.000) & 86.25 (0.29) & 12.01 (0.04)   \\ \cmidrule{2-7}
                                   & Reg.~with log-transform   & 22.41 (0.07) & $-4.56$ (0.01) & 0.015 (0.000) & 96.02 (0.75) & 11.88 (0.03)   \\ \cmidrule{2-7}
                                   & Gaussian copula reg. & 22.11 (0.07) & $-4.46$ (0.00) & 0.015 (0.000) & 84.79 (0.27) & 11.78 (0.03)   \\ \cmidrule{2-7}
                                   & Vine copula reg.     & 22.43 (0.07) & $-4.44$ (0.00) & 0.016 (0.000) & 82.42 (0.26) & 11.91 (0.04)  \\ \midrule
\multirow{3}{*}{3}                 & Linear reg.          & 1.22 (0.00)  & $-1.62$ (0.00) & 0.251 (0.001) & 4.80 (0.02) & 0.67 (0.00)   \\ \cmidrule{2-7}
                                   & Reg.~with log-transform   & 1.22 (0.00) & $-1.57$ (0.00) & 0.270 (0.001) & 4.73 (0.02) & 0.64 (0.00)   \\ \cmidrule{2-7}
                                   & Quadratic reg.       & 1.13 (0.00)  & $-1.54$ (0.00) & 0.275 (0.001) & 4.42 (0.01) & 0.62 (0.00)   \\ \cmidrule{2-7}
                                   & Gaussian copula reg. & 1.21 (0.00)  & $-1.56$ (0.00) & 0.273 (0.001) & 4.68 (0.02) & 0.64 (0.00)   \\ \cmidrule{2-7}
                                   & Vine copula reg.     & 1.19 (0.00)  & $-1.50$ (0.00) & 0.290 (0.001) & 4.35 (0.01) & 0.63 (0.00)   \\ \bottomrule
\end{tabular}
\caption{
Simulation results for four explanatory variables.
The table shows the root-mean-square error (RMSE), logarithmic score (LogS), 
quadratic score (QS), interval score (IS), and integrated Brier score (IBS) in different simulation cases. 
The arrows in the header indicate that lower RMSE, IS, and IBS; 
and higher LogS and QS are better.
The numbers in parentheses are the corresponding standard errors.}
\label{tab: simulation_comparison_dim5}
\end{table}

\section{Application}
\label{sec:application}

\subsection{Abalone data set}
\begin{figure}
    \centering
    \includegraphics[width=0.7\textwidth]{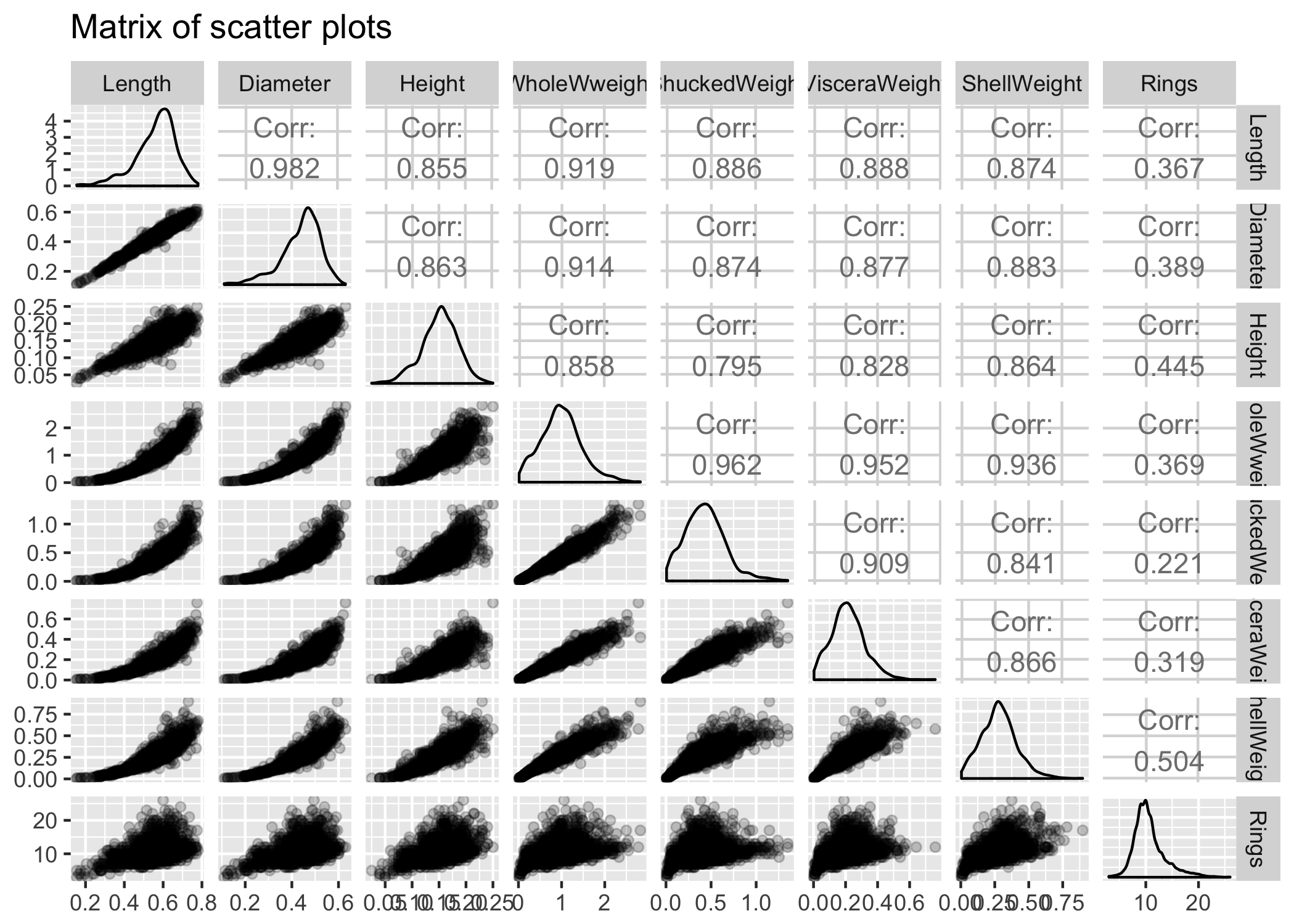}
    \caption{Pairwise scatter plots of the Abalone dataset.}
    \label{fig:abalone_pairs_plot}  
\end{figure}

In this section, we apply the vine copula regression method on a real data set: the Abalone data set \citep{Lichman:2013}.
The data set comes from an original (non-machine-learning) study~\citep{nash1994population}.
It has 4177 cases, and the goal is to predict the age of abalone from physical measurements; the names of these measurements are in \autoref{fig:abalone_pairs_plot}. 
The age of abalone is determined by counting the number of rings ({\tt Rings}) through a microscope, and this is a time-consuming task. Other physical measurements that are easier to obtain, are used to predict the age.
{\tt Rings} can be regarded either as a continuous variable or an ordinal one. Thus the problem can be either a regression or a classification problem. 
We focus on the subset of 1526 male samples (with two outliers removed).
\autoref{fig:abalone_pairs_plot} shows the pairwise scatter plots, marginal density functions and pairwise correlation coefficients.
There is clear non-linearity and heteroscedasticity among the pairs of variables.
We discuss the regression problem in \autoref{sec: application_regression}, and \autoref{sec: application_classification} shows the results for the classification problem.

\subsection{Regression}
\label{sec: application_regression}

In this section, we compare the performance of vine copula and linear regression methods.
Three vine regressions are considered:
\begin{itemize}
    \item R-vine copula regression: the proposed method with the candidate bivariate copula families;
    \item Gaussian copula regression with R-vine partial correlation parametrization: the proposed method with the bivariate Gaussian copulas only;
    \item D-vine copula regression: \citet{Kraus20171} with the candidate bivariate copula families. 
\end{itemize}
The candidate bivariate copulas include Student-$t$, MTCJ, Gumbel, Frank, Joe, BB1, BB6, BB7, BB8, and the corresponding survival and reflected copulas.

We perform 100 trials of 5-fold cross validation.
Vine copula regressions and linear regression are fitted using the training set, and the test set is used for performance evaluation.
All the univariate margins are fitted by skew-normal distributions.
The conditional mean and 95\% prediction interval are obtained for all 
models.
For copula regressions, the upper and lower bounds of the 95\% prediction interval are the conditional 97.5\% and 2.5\% quantiles respectively.

We consider the out-of-sample performance measures used in \autoref{sec: Simulation_Study}: the root-mean-square error (RMSE), logarithmic
score (LS), quadratic score (QS), interval score (IS), and integrated Brier score (IBS).
\autoref{tab: abalone_regression_comparison} shows the average performance 
measures from the 100 trials of cross validation.
Compared with linear regression, our method has lower prediction errors, and 
better predictive scores.
The performance of the R-vine copula model is slightly better than 
the D-vine copula model, in terms of all five scores.
The vine array and bivariate copulas on the edges of the R-vine fitted on the full dataset
are shown in \autoref{tab:vine_array_bi_cop}.
A visualization of the vine array is in 
\ref{sec: viz_abalone_vine_array}.
Several of the copulas linking to the response variables in trees 2 to 7
represent weak negative dependence.

The fitted D-vine regression model has path
\texttt{Diameter}--\texttt{VisceraWeight}--\texttt{WholeWeight}--\texttt{ShuckedWeight}--\texttt{ShellWeight}--\texttt{Rings}
in the first level of the D-vine structure.

We have also conducted monotonicity checks of the predicted
conditional median based on the fitted R-vine model.  Four of the linking
copulas in trees 2 to 7 (last column of the right-hand side of 
\autoref{tab:vine_array_bi_cop}) represent conditional negative dependence
given the previously linked variables to the response variable. 
This means that the conditional median function is not always monotone
increasing in an explanatory variable when others are held fixed.
However, when all explanatory variables are increasing together (for
larger abalone), the conditional median is increasing.  This property
is similar to classical Gaussian regression with positive correlated
explanatory variables and existence of negative regression coefficients
because of some negative partial correlations.  Even with some negative
conditional dependence, there is overall better out-of-sample prediction
performance by keeping all of the explanatory variables in the model.

We also did some numerical checks on the conditional quantiles when
one explanatory variable becomes extreme and other variables are held fixed.
It appears that the behavior is close to asymptotically constant.  From
the linking copulas in \autoref{tab:vine_array_bi_cop}
and the results in \autoref{sec:theoretical_results}, we would not be
expecting asymptotic linear behavior (and this is reasonable from the
context of the variables).

\begin{table}
\footnotesize
\centering
\begin{tabular}{lrrrrr}
\toprule
Model & RMSE$\,\downarrow$  & LogS$\,\uparrow$ & QS$\,\uparrow$ & IS$\,\downarrow$ & IBS$\,\downarrow$ \\ 
\midrule
Linear reg.            & 2.272 & $-2.240$ & 0.138 & 8.909 & 1.232 \\
Gaussian copula reg.   & 2.287 & $-2.142$ & 0.152 & 8.276 & 1.208 \\
D-vine copula reg.     & 2.183 & $-2.064$ & 0.163 & 8.104 & 1.141\\
R-vine copula reg.     & 2.168 & $-2.057$ & 0.164 & 8.005 & 1.136 \\
\bottomrule           
\end{tabular}
\caption{
Comparison of the performance of vine copula regressions and linear regression.
The numbers are the average scores over 100 trials of 5-fold cross validation.
The scoring rules are defined in 
\autoref{sec: Simulation_Study}.
}
\label{tab: abalone_regression_comparison}
\end{table}

\begin{table}
\footnotesize
\centering
\tabcolsep=3pt
\begin{tabular}{rrrrrrrr}
\toprule
4 & 4 & 4 & 4 & 4 & 7 & 1 & 7\\ \midrule
  & 7 & 7 & 7 & 5 & 4 & 4 & 4\\ \midrule
  &   & 5 & 5 & 7 & 6 & 5 & 5\\ \midrule
  &   &   & 6 & 6 & 5 & 7 & 6\\ \midrule
  &   &   &   & 1 & 1 & 6 & 3\\ \midrule
  &   &   &   &   & 3 & 3 & 1\\ \midrule
  &   &   &   &   &   & 2 & 2\\ \midrule
  &   &   &   &   &   &   & 8\\ \bottomrule
\end{tabular}
\quad  
\begin{tabular}{cccccccc}
\toprule
- & BB6.s & BB6.s & BB6.s & BB1.s & Gumbel.s & BB6.s & Gumbel.s\\
\midrule
- & - & t & Joe.v & BB8.s & t & BB8.s & BB8.u\\
\midrule
- & - & - & t & Frank & Frank & BB8.v & BB8.u\\
\midrule
- & - & - & - & Frank & t & Frank & MTCJ.v\\
\midrule
- & - & - & - & - & t & Frank & t\\
\midrule
- & - & - & - & - & - & Gumbel & t\\
\midrule
- & - & - & - & - & - & - & Gumbel.u\\
\midrule
- & - & - & - & - & - & - & -\\
\bottomrule
\end{tabular}
\caption{
Vine array and bivariate copulas of the R-vine copula regression fitted on the full dataset.
The variables are (1) \texttt{Length}, (2) \texttt{Diameter}, (3) \texttt{Height}, (4)
\texttt{WholeWeight}, (5) \texttt{ShuckedWeight}, (6) \texttt{VisceraWeight}, (7) \texttt{ShellWeight},
(8) \texttt{Rings}.
A suffix of `s' represents survival version of the copula family
to get the opposite direction of joint tail asymmetry; `u' and `v' represent the copula family with reflection on the first and second
variable respectively to get negative dependence.
}
\label{tab:vine_array_bi_cop}
\end{table}

\begin{figure}
    \centering
    \includegraphics[width=0.6\textwidth]{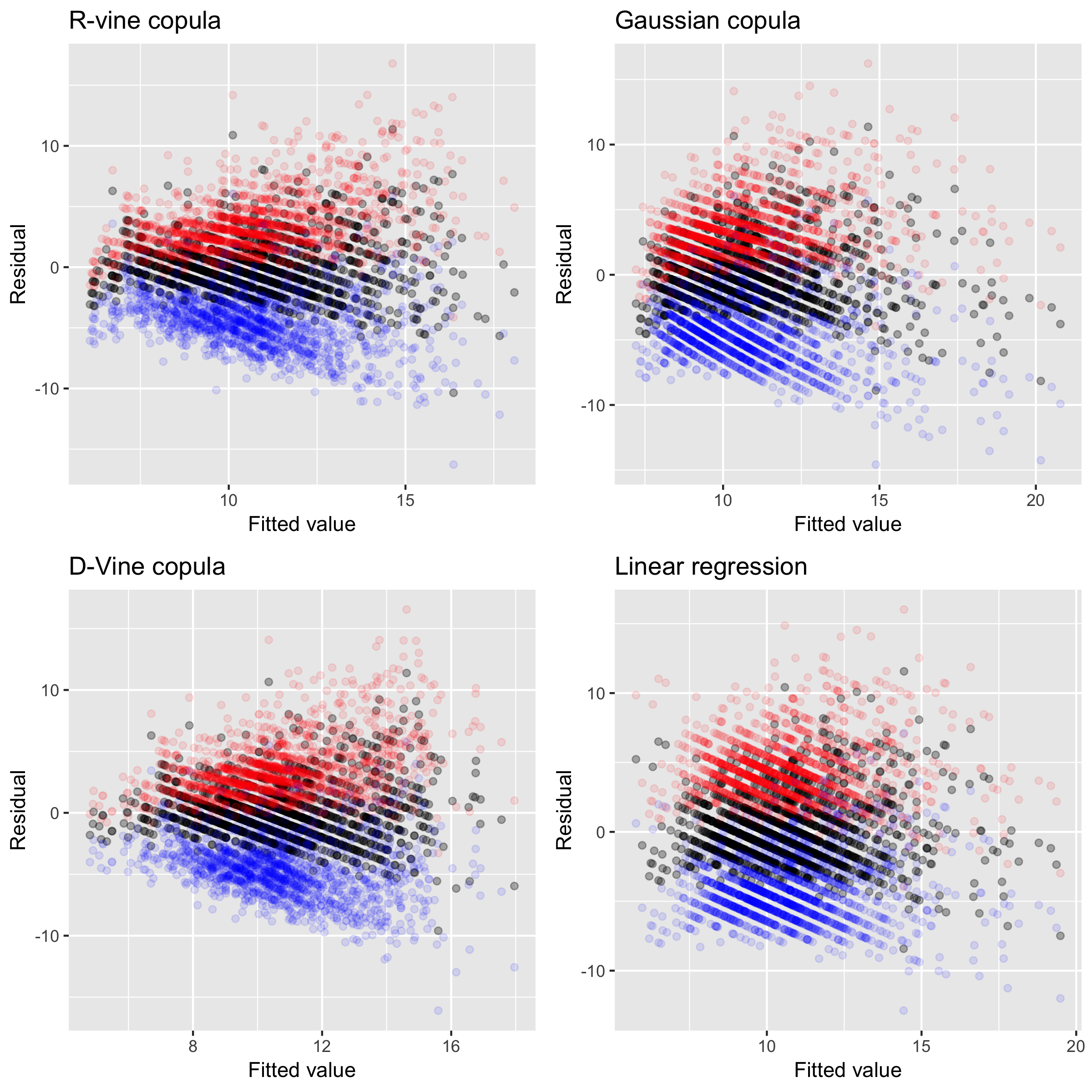}
    \caption{Residual vs. fitted value plots. The red and blue points correspond to the lower bound and upper bound of the prediction intervals.}
    \label{fig: abalone_residual_plot}   
\end{figure}

\autoref{fig: abalone_residual_plot} visualizes the prediction performance of the three methods on the full dataset.
The plots show the residuals against the fitted values on the test set, and the prediction intervals. 
Due to heteroscedasticity, there is more variation in residuals as fitted value increases.
However, linear regression fails to capture the heteroscedasticity and the prediction intervals are roughly of the same length.
Vine copula regression gives wider (narrower) prediction intervals when the fitted values are larger (smaller).
This illustrates the reason why our method overall has more precise prediction intervals.

In the supplementary material, 
further analysis is done to compare the
four methods to show where they differ the most in terms of point predictions.
The largest differences are when samples are near the upper boundary of the predictor space; 
that is, at least one of the predictor variables is above its 95th quantile.
This is an indication that R-vine copula and D-vine copula models are more flexible than 
Gaussian copula and linear regression models 
in handling tail behaviors.

\subsection{Classification}
\label{sec: application_classification}

The response variable \texttt{Rings} is an ordinal variable that ranges from 3 to 27. 
Therefore this is a multiclass classification problem.
Although our method can handle multiclass classification problems, we reduce it to a binary classification problem for easy comparison with commonly used methods, including logistic regression, support vector machine (SVM), and random forest (RF).
The sample median of \texttt{Rings} is 10; if a sample's \texttt{Rings} is greater than 10, we label it as `large', otherwise `small'.
All the predictor variables are fitted by skew-normal distributions, and we fit an empirical distribution to the response variable \texttt{Rings}.

The D-vine regression method \citep{Kraus20171} can only 
handle continuous variables and is not directly applicable to the 
classification problem.
In order to compare our method with the D-vine based method, we first treat the binary response variable as a continuous variable (0 and 1) and use the D-vine regression method \citep{Kraus20171} to find a D-vine structure or an ordering of variables. 
Then an R-vine regression model is fitted on that D-vine structure using our method.

\begin{figure}
    \centering
    \begin{subfigure}[b]{0.45\textwidth}
        \centering
        \includegraphics[width=\textwidth]{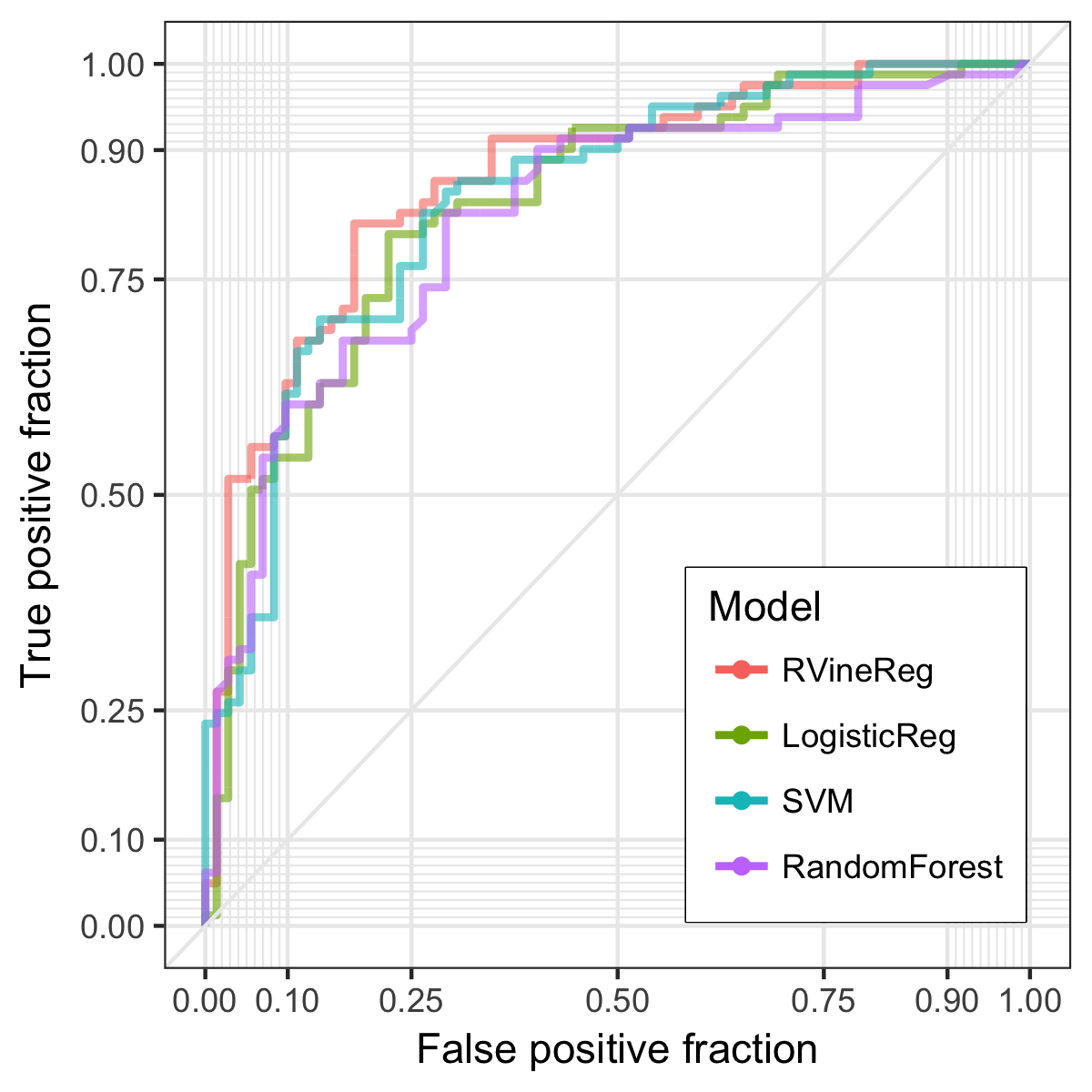}
        \caption{ROC curves of different binary classifiers. 
        The performance is evaluated on the test set. }
        \label{fig: abalone_roc}   
    \end{subfigure}%
    \qquad
    \begin{subfigure}[b]{0.45\textwidth}
        \centering
        \includegraphics[width=\textwidth]{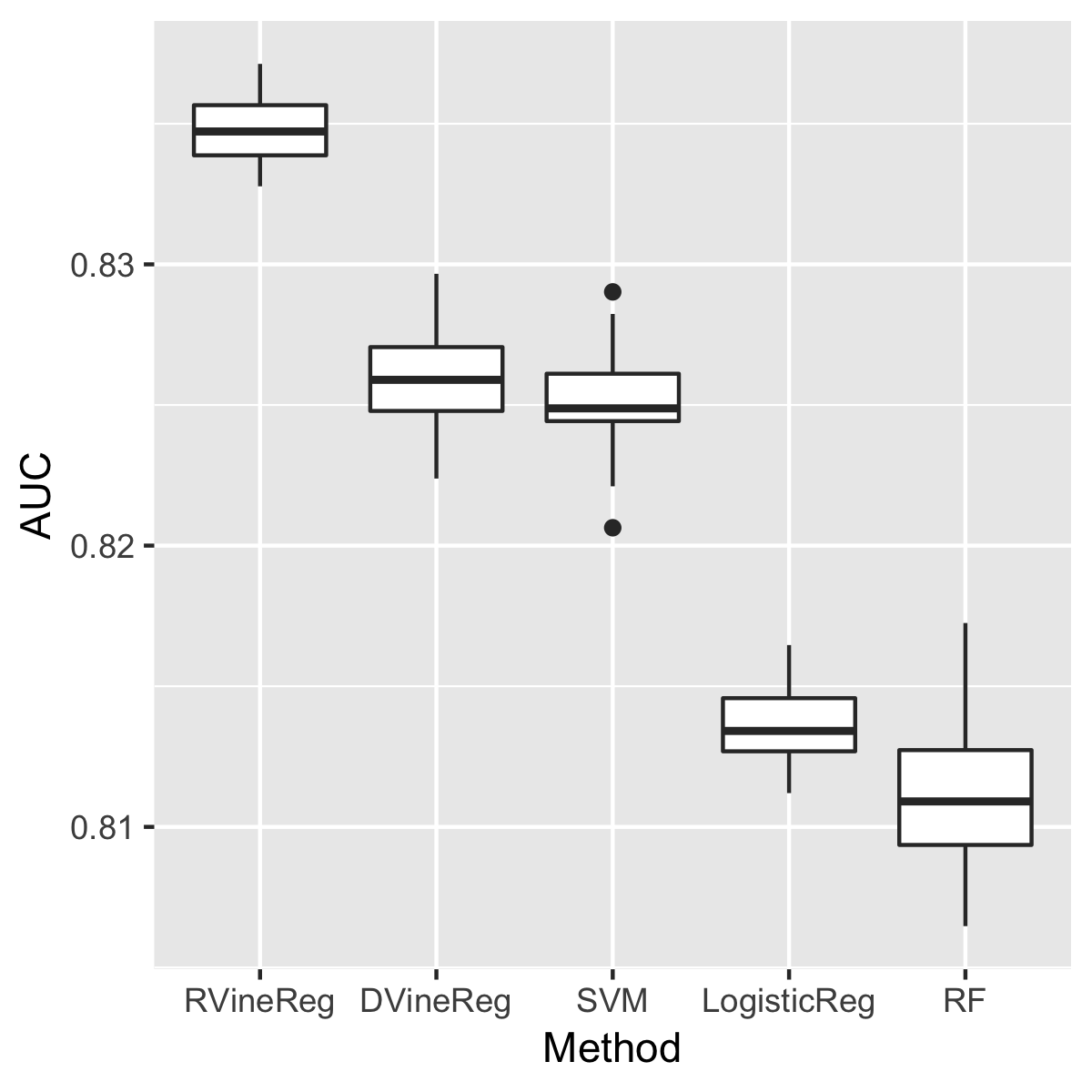}
        \caption{Box plot of the AUCs bases on 10-fold cross validations, repeated 20 times.}
        \label{fig: abalone_auc_boxplot}
    \end{subfigure}
    \caption{Comparison of the performance on the classification problem.}
\end{figure}

For binary classifiers, the performance can be demonstrated by a receiver operating characteristic (ROC) curve.
The curve is created by plotting the true positive rate against the false positive rate at various threshold settings.
The $(0,1)$ point corresponds to a perfect classification; a completely random guess would give a point along the diagonal line. 
An ROC curve is a two-dimensional depiction of classifier performance. To compare classifiers we may want to reduce ROC performance to a scalar value representing the expected performance. A common method is to calculate the area under the ROC curve, abbreviated AUC \citep{Fawcett:2006:IRA:1159473.1159475}.
The AUC can also be interpreted as the probability that a classifier will rank a randomly chosen positive instance higher than a randomly chosen negative one.
Therefore, larger AUC is better.
\autoref{fig: abalone_roc} shows sample ROC curves of different binary classifiers and the corresponding AUCs.

Repeated 10-fold cross validations with random partitions is used to assess the performance. 
In each pass, a 10-fold cross validation is performed and the average AUC is recorded.
\autoref{fig: abalone_auc_boxplot} shows a box plot of the average AUCs.
The performance of vine copula regression is marginally better than the other methods.
The average AUCs are: RVineReg = 0.835, DVineReg = 0.826, SVM = 0.825, LogisticReg = 0.814, RF = 0.811.

\section{Conclusion}
\label{sec:conclusion}

Our vine copula regression method uses R-vines and can fit mixed continuous and ordinal variables.
The prediction algorithm can efficiently compute the conditional distribution given a fitted vine copula, without marginalizing the conditioning variables.
The performance of the proposed method is evaluated on simulated data sets and the Abalone data set.
The heteroscedasticity in the data is better captured by vine copula regression than the standard regression methods.

One potential drawback of the proposed method is the computational cost for high-dimensional data, especially when the dimensionality is greater than the sample size.
This paper is more of a proof of concept of using R-vine copula models for regression and classification problems. Therefore, we evaluate the performance of the proposed methods on classical cases and compare with models such as linear regressions.
Another drawback is the constraint on the vine structure such that the response variable is always a leaf node at each level. 
This constraint greatly reduces the computational complexity; without it, numerical integration would be required to compute the conditional CDF.

To relate how choices of bivariate copula families in the vine can affect prediction and to provide guidelines on bivariate copula families to consider, we give a theoretical analysis of the asymptotic shape of conditional quantile functions. For bivariate copulas, the conditional quantile function of the response variable could be asymptotically linear, sublinear, or constant with respect to the explanatory variable. It turns out the asymptotic conditional distribution can be quite complex for trivariate and higher-dimensional cases, and there are counter-intuitive examples. In practice, we recommend remarkable plots of conditional quantile functions of the fitted vine copula to assess if the monotonicity properties are reasonable.

One possible future research direction is the extension of the proposed regression method for survival outcomes with censored data. For example, \citet{emura2018personalized} use bivariate copulas to predict time-to-death given time-to-cancer progression; \citet{barthel2018vine} apply vine copulas to multivariate right-censored event time data. They apply copulas to the joint survival function instead of the joint CDF to deal with right-censoring. These types of applications would require more numerical integration methods.

Another research direction is to handle variable selection and reduction when there are many explanatory variables, some of which might form clusters with strong dependence.
Traditional variable selection methods for regression can also be applied, for example, the forward selection approach.
Moreover, recent papers proposed methods for learning sparse vine copula models \citep{nagler2019model,muller2019dependence}, which can be potentially used as a variable selection method for copula regression.

\section*{Acknowledgments}

This research has been supported by an NSERC Discovery Grant 8698, and a Collaborative Research Team grant for the project: \textit{Copula Dependence Modeling: Theory and Applications} of the Canadian Statistical Sciences Institute.
We are grateful to the referees and associate editor for comments leading
to an improved presentation.

\bibliographystyle{model5-names}
\bibliography{biblio}

\appendix

\section{Vine array representation}
\label{sec:vine_array_representation}

An R-vine can be represented by the edge sets at each level $E(T_{\ell})$, or equivalently by a graph, such as \autoref{fig: 2-trunc_vine_example}. 
But those representations are not convenient; we need a more compact way to represent vine models.
A vine array $A=(a_{jk})$ for a regular vine $\mathcal{V} = (T_1, \ldots, T_{d-1})$ on $d$ elements is a $d \times d$ upper triangular matrix. 
There is an ordering of the variable indexes along the diagonal,
and row $\ell$, column $j$ shows the variable $a_{\ell j}$
is connected to the variable $a_{\ell\ell}$ in tree $\ell$, conditioning
on variables $a_{1j},\ldots,a_{\ell-1,j}$.
That is,
the first $\ell$ rows of $A$ and the diagonal elements encode the $l$-th tree $T_{\ell}$, such that $[a_{\ell j}, a_{jj} | a_{1j}, \ldots , a_{\ell-1, j}] \in E(T_{\ell})$, for $\ell + 1 \leq j \leq d$.
For example, the vine array $A_1$ represents the R-vine in \autoref{fig: 2-trunc_vine_example}.
The edges of $T_1$ include $[a_{12}, a_{22}] = [23]$, $[a_{13}, a_{33}] = [24]$, $[a_{14}, a_{44}] = [21]$, $[a_{15}, a_{55}] = [35]$.    
The edges of $T_2$ include $[a_{23}, a_{33} | a_{13}] = [34|2]$, $[a_{24}, a_{44} | a_{14}] = [31|2]$,  $[a_{25}$, $a_{55} | a_{15}] = [25|3]$.
\begin{equation*}
A_1 = 
 \begin{pmatrix}
  2 & 2 & 2 & 2 & 3 \\
    & 3 & 3 & 3 & 2 \\
    &   & 4 & 4 & 4 \\
    &   &   & 1 & 1 \\
    &   &   &   & 5 
 \end{pmatrix},
\quad
 A_2 = 
 \begin{pmatrix}
  2 & 2 & 2 & 3 & 2 \\
    & 3 & 3 & 2 & 3 \\
    &   & 4 & 4 & 4 \\
    &   &   & 5 & 5 \\
    &   &   &   & 1 
 \end{pmatrix}.
\end{equation*}
Note that a valid vine array represent a unique R-vine. 
However, a R-vine may have multiple vine array representations. 
For example, $A_1$ and $A_2$ encode exactly the same R-vine.
In real applications, the variables are labeled arbitrarily.
We can define a permutation of the variables so that the diagonal elements are $(1, 2, \ldots, d)$.
Therefore, \autoref{alg: rvinepcond mix} only applies to vine arrays with ordered diagonal elements.

\section{Visualization of the R-vine array in \autoref{tab:vine_array_bi_cop}}
\label{sec: viz_abalone_vine_array}

\begin{figure}
\centering
\begin{subfigure}[b]{0.33\textwidth}
    \centering
    \includegraphics[width=\textwidth]{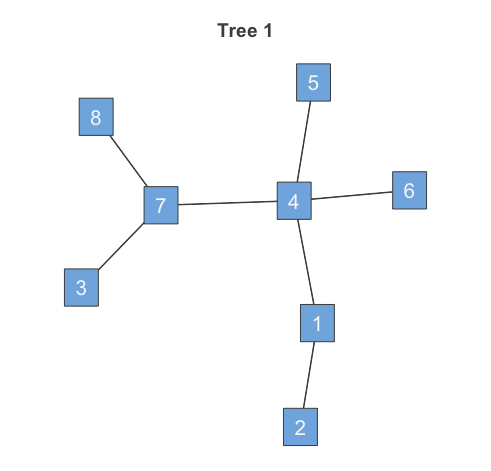}
\end{subfigure}%
\begin{subfigure}[b]{0.33\textwidth}
    \centering
    \includegraphics[width=\textwidth]{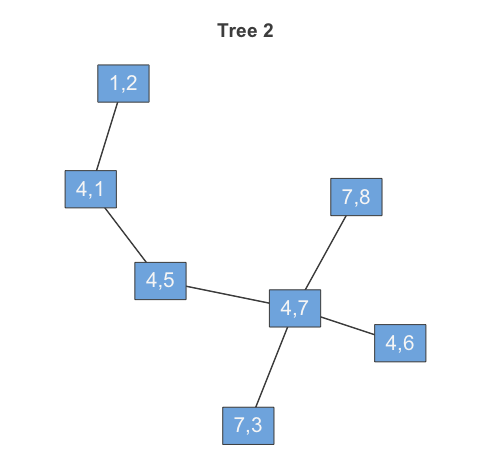}
\end{subfigure}
\begin{subfigure}[b]{0.33\textwidth}
    \centering
    \includegraphics[width=\textwidth]{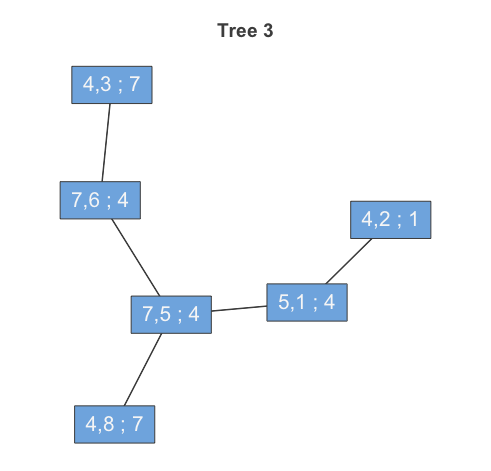}
\end{subfigure}
\\
\begin{subfigure}[b]{0.33\textwidth}
    \centering
    \includegraphics[width=\textwidth]{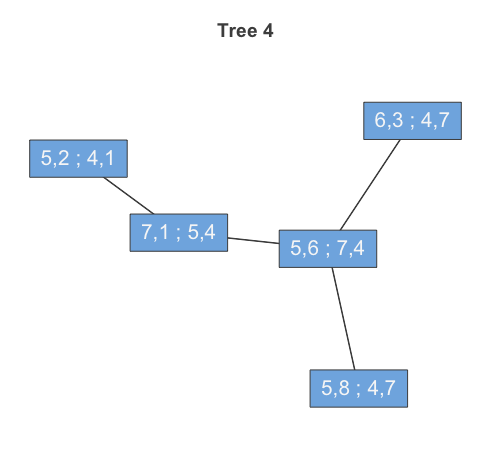}
\end{subfigure}%
\begin{subfigure}[b]{0.33\textwidth}
    \centering
    \includegraphics[width=\textwidth]{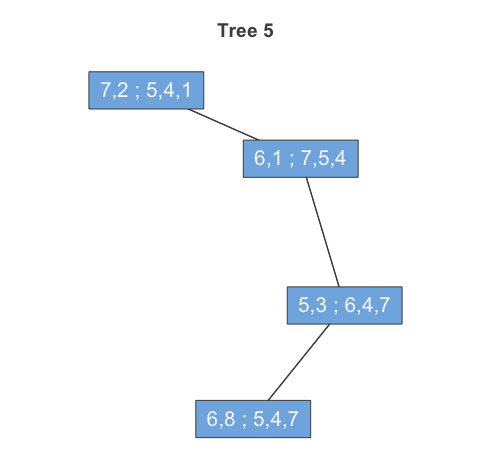}
\end{subfigure}
\begin{subfigure}[b]{0.33\textwidth}
    \centering
    \includegraphics[width=\textwidth]{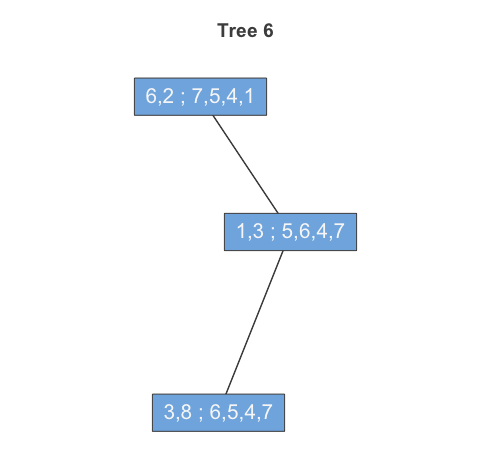}
\end{subfigure}
\\
\begin{subfigure}[b]{0.33\textwidth}
    \centering
    \includegraphics[width=\textwidth]{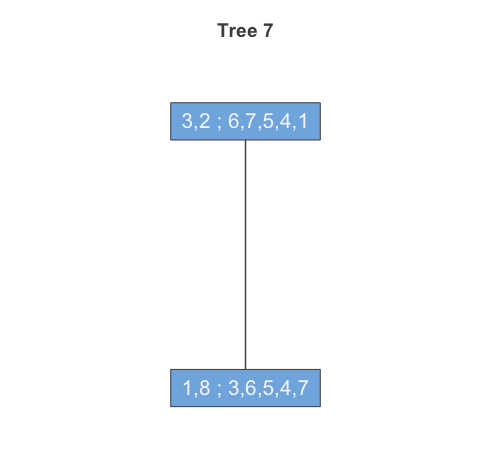}
\end{subfigure}
\caption{Visualization of the R-vine array in \autoref{tab:vine_array_bi_cop}.}
\label{fig: abalone_Rvine}
\end{figure}

\section{Contents of supplementary materials}

The contents of the supplementary materials are organized as follows.

\dirtree{%
.1 supplementary/.
.2 abalone.
.3 abalone.data.
.3 abalone\_classification.R.
.3 abalone\_regression.R.
.2 copreg\_0.190323.tar.gz.
.2 README.txt.
.2 simulation.
.3 linear\_hetero.R.
.3 linear\_homo.R.
.3 nonlinear\_hetero.R.
.3 simulation.R.
.2 supplementary.pdf.
}

\begin{itemize}
\item {\tt abalone/abalone.data}: The raw data file for the dataset used in Section~\ref{sec:application}.
\item {\tt abalone/abalone\_classification.R}: R code for data analysis in Section~\ref{sec: application_classification}.
\item {\tt abalone/abalone\_regression.R}: R code for data analysis in Section~\ref{sec: application_regression}.
\item {\tt copreg\_0.190323.tar.gz}: R package for R-vine copula regression methods.
\item {\tt README.txt}: README file.
\item {\tt simulation/linear\_hetero.R}: R code for data analysis in Section~\ref{sec: Simulation_Study}, the linear and heteroscedastic case.
\item {\tt simulation/linear\_homo.R}: R code for data analysis in Section~\ref{sec: Simulation_Study}, the linear and homoscedastic case.
\item {\tt simulation/nonlinear\_hetero.R}: R code for data analysis in Section~\ref{sec: Simulation_Study}, the Non-linear and heteroscedastic case.
\item {\tt simulation/simulation.R}: R functions related to the simulation study.
\item {\tt supplementary.pdf}: Theoretical results on shapes of conditional quantile functions.
\end{itemize}

\end{document}